\documentclass[nohyper,12pt,letterpaper]{JHEP3}
\usepackage{amsmath}
\usepackage{amsthm}
\usepackage{amsfonts}

\DeclareMathOperator{\tr}{tr} 
\DeclareMathOperator{\Imx}{Im} 
\DeclareMathOperator{\Rex}{Re} 

\DeclareMathOperator{\Spin}{Spin} 

\date{\today}

\title{Selfduality and Chern--Simons Theory}

\author{
Ori J. Ganor and Yoon Pyo Hong
\\
\\
Department of Physics,
University of California, Berkeley, CA 94720
\\
Emails: 
\email{origa@socrates.berkeley.edu,yph@berkeley.edu}\\
}

\abstract{
We propose a relation between the operator of S-duality
(of ${\cal N}=4$ super Yang--Mills theory in 3+1D)
and a topological theory in one dimension lower.
We construct the topological theory by
compactifying ${\cal N}=4$ super Yang--Mills
on $S^1$ with an S-duality and R-symmetry twist.
The S-duality twist requires a selfdual coupling constant.
We argue that for a sufficiently low rank of the gauge group
the three-dimensional low-energy description is a topological
theory, which we conjecture to be a pure Chern--Simons theory.
This conjecture implies a connection
between the action of mirror
symmetry on the sigma-model with Hitchin's moduli space as target space
and geometric quantization of the moduli space of flat connections
on a Riemann surface.
}

\received{December 5, 2008}

\keywords{S-Duality, Chern--Simons, Supersymmetry,
Flat Connections, Hitchin Space,
T-duality, Mirror symmetry, Geometric quantization}
\preprint{
UCB-PTH-08/71
}

\begin{document}

\def\be{\begin{equation}}
\def\ee{\end{equation}}
\def\bear{\begin{eqnarray}}
\def\eear{\end{eqnarray}}
\def\nn{\nonumber}

\newcommand\belabel[1]{\begin{equation}\label{#1}}
\newcommand\bearlabel[1]{\begin{eqnarray}\label{#1}}


\newcommand\SUSY[1]{{${\cal N}={#1}$}}  
\newcommand\px[1]{{\partial_{#1}}}
\newcommand\qx[1]{{\partial^{#1}}}
\newcommand\bpx[1]{{\overline{\partial}_{#1}}}
\newcommand\bqx[1]{{\overline{\partial}^{#1}}}

\newcommand\ppx[1]{{\frac{\partial}{\partial {#1}}}}
\newcommand\pspxs[1]{{\frac{\partial^2}{\partial {#1}^2}}}
\newcommand\pypx[2]{{\frac{\partial {#1}}{\partial {#2}}}}

\def\a{\alpha}
\def\b{\beta}
\def\g{\gamma}

\def\dta{{\dot{\a}}}
\def\dtb{{\dot{\b}}}
\def\dtg{{\dot{\g}}}
\def\dtd{{\dot{\d}}}

\def\Mf{{M}}
\def\MXf{{X}}
\def\gYM{g_{\text{YM}}}
\def\Ham{{H}} 

\def\sdA{{\widetilde{A}}} 
\def\sdPhi{{\widetilde{\Phi}}} 
\def\sdpsi{{\widetilde{\psi}}} 
\def\sdtau{{\widetilde{\tau}}} 
\def\sdHam{{\widetilde{\Ham}}} 

\def\sdtheta{{\widetilde{\theta}}} 
\def\sdgYM{\tilde{g}_{\text{YM}}}

\def\th{{\theta}}
\def\lam{{\lambda}}
\def\bpsi{{\overline{\psi}}}
\def\bchi{{\overline{\chi}}} 

\def\oa{{\overline{a}}} 

\def\bth{{\overline{\theta}}}
\def\blam{{\overline{\lambda}}}
\def\bsig{{\overline{\sigma}}}
\def\Dslash{{\relax{\not\kern-.18em \partial}}} 
\def\Spin{{{\mbox{\rm Spin}}}} 
\def\SL{{{\mbox{\rm SL}}}} 
\def\GL{{{\mbox{\rm GL}}}} 

\newcommand{\field}[1]{\mathbb{#1}}
\newcommand{\ring}[1]{\mathbb{#1}}
\newcommand{\C}{\field{C}}
\newcommand{\R}{\field{R}}
\newcommand{\Z}{\ring{Z}}
\newcommand{\N}{\ring{N}}

\newcommand{\CP}{\mathbb{P}} 


\providecommand{\abs}[1]{{\lvert#1\rvert}}
\providecommand{\norm}[1]{{\lVert#1\rVert}}
\providecommand{\divides}{{\vert}}
\providecommand{\suchthat}{{:\quad}}


\newcommand\rep[1]{{\bf {#1}}} 


\def\Horava{{Ho\v{r}ava\ }}
\def\Cech{{\v{C}ech\ }}


\numberwithin{equation}{section}

\newcommand{\secref}[1]{\S\ref{#1}}

\newcommand{\figref}[1]{Figure~\ref{#1}}
\newcommand{\appref}[1]{Appendix~\ref{#1}}
\newcommand{\tabref}[1]{Table~\ref{#1}}


\def\cD{{\cal D}} 
\def\utr{{\mbox{tr}}}

\def\tA{{\widetilde{A}}} 
\def\tW{{\widetilde{W}}} 
\def\tE{{\widetilde{E}}} 
\def\tB{{\widetilde{B}}} 

\def\Mst{M_{\text{st}}} 
\def\gst{g_{\text{st}}} 
\def\lst{\ell_{\text{st}}} 
\def\lp{\ell_{\text{P}}} 

\def\apr{{\a'}}

\def\Car{{\cal H}} 

\def\SKer{{\cal S}} 
\def\TKer{{\cal T}} 
\def\ASKer{{\cal S}_A} 
\def\Link{{\cal L}} 

\def\lV{{V}} 
\def\sdlV{{\widetilde{\lV}}} 

\def\IKer{{I}} 
\def\gtw{{\gamma}} 
\def\ftw{{\varphi}} 
\def\Area{{\cal A}} 
\def\XS{{X}} 

\def\wvF{{\Psi}} 
\def\sdwvF{{\widetilde{\Psi}}} 

\def\sdT{{\widetilde{T}}} 

\def\cgC{{\cal C}} 

\newcommand\bra[1]{{\langle{#1}\rvert}} 
\newcommand\ket[1]{{\lvert{#1}\rangle}} 
\def\Op{{\cal O}} 

\def\xP{{\Omega}} 
\def\sdxP{{\widetilde{\Omega}}} 

\def\xa{{\mathbf{a}}} 
\def\xb{{\mathbf{b}}} 
\def\xc{{\mathbf{c}}} 
\def\xd{{\mathbf{d}}} 

\def\zs{{\mathbf{s}}} 

\def\xA{{\mathbf{A}}} 
\def\xB{{\mathbf{B}}} 
\def\xC{{\mathbf{C}}} 
\def\xD{{\mathbf{D}}} 

\def\zt{{\mathbf{t}}} 

\def\bQ{{\overline{Q}}} 

\def\OpS{{\widehat{{\cal S}}}} 

\def\AcV{{\cal I}} 
\def\LagJ{{\omega}} 
\def\fx{{q}} 
\def\lvk{{k}} 
\def\pht{{\upsilon}} 
\def\Pauli{\mathfrak{\sigma}} 
\def\CSA{\mathfrak{h}} 

\def\tdlphi{\tilde{\phi}} 
\def\tdlPhi{\tilde{\Phi}} 
\def\tdlw{\tilde{w}} 
\def\tdlPsi{\widetilde{\Psi}} 

\def\MH{{{\cal M}_H}} 
\def\MFC{{{\cal M}_{\text{fc}}}} 
\def\dimT{{d}} 
\def\zsGroup{{\mathbb{S}}} 

\def\Id{{\mathbf{I}}} 

\def\vE{{\vec{E}}} 
\def\vB{{\vec{B}}} 
\def\Zphi{{Z}} 
\def\ef{\mathbf{e}} 
\def\mf{\mathbf{m}} 

\def\ActS{{S}} 
\def\XPhi{{\cal Z}} 
\def\wssig{{\tilde{\sigma}}} 
\def\wstau{{\tilde{\tau}}} 

\def\TgSp{{Y}} 
\def\Lbd{{\cal L}} 
\def\bj{{\overline{j}}} 
\def\PatchU{{\cal U}} 

\def\RSC{{{\cal C}}} 
\def\LoopC{{\mathbf{C}}} 

\def\bz{{\overline{z}}} 
\def\bphi{{\overline{\phi}}} 
\def\CanK{{\cal K}} 
\def\bCanK{{\overline{\CanK}}} 
\def\twsu{{\mathbf{T}}} 
\def\twx{{\varrho}} 

\def\bXPhi{{\overline{\XPhi}}}
\def\brho{{\overline{\rho}}}

\def\ZXPhi{{Z}} 
\def\bZXPhi{{\overline{\ZXPhi}}} 

\def\Sp{{{\mbox{\rm Sp}}}} 
\def\PF{{\cal Z}} 
\def\aRoot{{\alpha}} 

\def\sdLoopC{{\widetilde{\LoopC}}} 
\def\Linking{{L}} 

\def\btau{{\overline{\tau}}} 
\def\pz{{q}}


\section{Introduction}
\label{sec:intro}

In the last 14 years there has been a lot of progress
on the subject of S-duality of \SUSY{4}
super Yang--Mills theory (SYM).
The conjecture \cite{Montonen:1977sn,Goddard:1976qe,Osborn:1979tq}
has passed many elaborate tests,
including those in the following list:
the number of BPS dyons obeys S-duality \cite{Sen:1994yi},
the partition function of a topologically twisted theory
is S-dual \cite{Vafa:1994tf},
and after a supersymmetric compactification
on a Riemann surface
S-duality reduces to mirror symmetry
in the low-energy limit \cite{Bershadsky:1995vm,Harvey:1995tg}.
The moduli space of mass-deformed \SUSY{4}
has been established to obey S-duality \cite{Seiberg:1994aj},
the action of S-duality on operators has been deduced
for many local operators \cite{Intriligator:1998ig},
as well as Wilson loops and `t Hooft loops
\cite{Kapustin:2005py}, 
it was demonstrated that low-energy states of
the toroidally compactified theory
also obey S-duality \cite{Henningson:2008ri},
and a comprehensive framework for
determining the action of S-duality on BPS boundary conditions
\cite{Gaiotto:2008sa} has been developed in \cite{Gaiotto:2008ak}.
S-duality also fits nicely within the framework of
the AdS/CFT correspondence \cite{Aharony:1999ti}.\par
New insights on S-duality emerged from various
sources: Witten's conjecture \cite{Witten:1995zh}
that \SUSY{4} SYM is the low-energy limit of a $T^2$
compactification of a
six-dimensional conformal field theory,
the $(2,0)$-theory, gave rise to a
geometrical realization of S-duality with simply-laced groups,
and this has been generalized to other groups as well
\cite{Vafa:1997mh};
M(atrix) theory \cite{Banks:1996vh}
and its application to the $(2,0)$-theory
\cite{Aharony:1997th,Aharony:1997an} therefore
led to a conjectured S-duality invariant
formulation of \SUSY{4} SYM \cite{Ganor:1997jx}.
More recently, new insight has emerged
about the connection between S-duality and
the geometric Langlands program \cite{Kapustin:2006pk}.

But S-duality still remains a mystery.
What is required is an operator $\SKer$ that
transforms a state in the Hilbert space of \SUSY{4}
SYM to a dual state in the Hilbert space of the dual theory.
(The approach of studying the Hilbert space
of a gauge theory was a fruitful one in three-dimensions
\cite{Karabali:1995ps,Leigh:2004ja}
and also an interesting direction
in four-dimensions \cite{Freidel:2008pw}.)
The operator $\SKer$ is directly related to the
action of S-duality on boundary conditions, as defined
in \cite{Gaiotto:2008ak}.\footnote{We are grateful to
E.~Witten for explaining this point to us.}
In this paper we will attempt to gather new clues about the
nature of the operator $\SKer$.
We will argue that this operator is related to
a three-dimensional nonlocal topological field theory.
We will also argue that under certain restrictions
a {\it local} topological field theory emerges.
The physical questions that define observables
in this local theory are as follows.
Suppose we compactify \SUSY{4} SYM on $S^1$
(parameterized by a periodic coordinate $0\le x_3<2\pi R$),
which we shall treat as (Euclidean) time,
but instead of setting up periodic boundary conditions whereby
the quantum states of the system
at $x_3=0$ and $x_3=2\pi R$ are required to agree, we instead
require that the state at $x_3=2\pi R$ is the {\it S-dual}
of the state at $x_3=0.$ This is an ``S-duality twist,''
which is possible at certain special selfdual
values of the coupling constant.
Our question now is: {\it what is the low-energy
effective three-dimensional description of this theory?}
With a few additional modifications and restrictions,
we will argue that this is a topological theory,
and we propose that it is a Chern--Simons theory.
Observables can then be constructed from
Wilson lines that are at a constant $x_3$
and we expect them to reduce to Wilson lines
in Chern--Simons theory. The expectation values of these
observables are related to knot invariants \cite{Witten:1988hf}.
[For the rest of this paper, unless otherwise noted,
$x_3$ will be understood as a spatial 
(as opposed temporal) direction.]
\\[0pt]
\indent
The paper is organized as follows.
In \secref{sec:problem} we construct the circle compactification
with the S-duality twist. We also introduce an R-symmetry
twist in order to preserve supersymmetry and reduce the number
of zero modes, 
and we add restrictions on the rank of the gauge group 
in order to eliminate all zero modes.
In \secref{sec:Kernel} we study the S-duality operator
from the Hilbert space perspective, and argue that
a topological three-dimensional action can be constructed from it.
In \secref{sec:Ttwist} we take a detour to study a related
problem in two-dimensional conformal field theory.
We discuss a $\sigma$-model with target space $T^d$
at a selfdual point in moduli space,
and show that compactification on $S^1$ with a T-duality
twist reduces in low-energy to a 0+1D topological theory
which can be identified with geometric quantization of
the target space. 
We then discuss a possible generalization of this result to
\SUSY{(2,2)} supersymmetric $\sigma$-models
which are selfdual under mirror symmetry.
In \secref{sec:Abelian} we return to four-dimensional
\SUSY{4} SYM and describe the abelian case where exact
results are well-known.
In \secref{sec:Nonabelian} we turn to the S-duality
and R-symmetry twisted circle compactification of
nonabelian \SUSY{4} SYM. We propose that the low-energy
topological theory is pure Chern--Simons theory,
and outline a test of this conjecture whereby we
compactify on a Riemann surface of genus $2$ and count
the number of vacua. Because of several unknown 
(at least to us) signs in the action of S-duality,
we only get partial results.
We conclude in \secref{sec:Discussion} with a discussion
and suggestion for further explorations.
\nopagebreak

\section{The problem}
\label{sec:problem}
Four-dimensional
\SUSY{4} super Yang--Mills theory with gauge group $U(n)$
is believed to possess $\SL(2,\Z)$-duality.
The complex coupling constant
$$
\tau\equiv \frac{4\pi i}{\gYM^2} + \frac{\theta}{2\pi}
$$
transforms under an element
$$
\begin{pmatrix} \xa & \xb \\ \xc & \xd \\ \end{pmatrix}
\in\SL(2,\Z)
$$
as
$$
\tau\rightarrow \frac{\xa\tau+\xb}{\xc\tau+\xd}\,.
$$

In this paper we are concentrating on {\it selfduality},
which occurs at values of $\tau$
for which there exists an element $\zs\in\SL(2,\Z),$
other than
$
\Id\equiv\left(\begin{array}{rr}
1 & 0 \\ 0 & 1 \\ \end{array}\right)
$
or $-\Id,$
that leaves $\tau$ invariant.
A selfdual theory can be compactified on an $S^1$
with an $\zs$-twist.
We are interested in the 2+1D
low-energy limit of this setting.
We will add a few more ingredients and restrictions in order
to preserve supersymmetry and to eliminate zero modes,
and we will argue that the resulting 2+1D theory
is a {\it topological field theory.}
The problem is:
{\it what is that topological field theory?}
We now turn to the details.

\subsection{S-duality twist}
\label{subsec:Stwist}
Up to $\SL(2,\Z)$-conjugation,
selfduality occurs in the following two cases:
\begin{enumerate}
\item
$\tau=i$ is fixed by the $\Z_4$ subgroup of $\SL(2,\Z)$
generated by
\be\label{eqn:zspr}
\zs'\equiv\left(\begin{array}{rr}
 0 & -1 \\ 1 & 0 \\ \end{array}\right)
\in\SL(2,\Z);
\ee
\item
$\tau=e^{\pi i/3}$ is fixed by the $\Z_6$
subgroup generated by
\be\label{eqn:zsprpr}
\zs''\equiv\left(
\begin{array}{rr} 1 & -1 \\ 1 & 0 \\ \end{array}
\right)\in\SL(2,\Z)\,.
\ee
\end{enumerate}
For a selfdual value $\tau$, any duality transformation $\zs$
that fixes it is a symmetry.
The elements $\zs'$ or $\zs''$ above
generate subgroups of $\SL(2,\Z)$
(either $\Z_4$ or $\Z_6$)
which are discrete gauge symmetries.
For a given $\tau$, we will denote the subgroup
of $\SL(2,\Z)$ that fixes it by $\zsGroup_\tau.$
On a manifold $\MXf$
with nontrivial first homotopy
group $\pi_1(\MXf)$, we can formulate the theory with
boundary conditions
that are twisted by elements of $\zsGroup_\tau$ along nontrivial loops.
The complete set of choices for such boundary conditions
is given by the set of homomorphisms
(maps that preserve the group structure)
from $\pi_1(\MXf)$ to $\zsGroup_\tau.$
In this paper, we will only study the case $\MXf=S^1\times\Mf_3$,
where $\Mf_3$ is some 3-manifold, 
with an $\SL(2,\Z)$-twist only along $S^1$.

Ignoring $\Mf_3$ for the moment,
we have 3+1D \SUSY{4} $U(n)$
SYM compactified on $S^1$ with an $\zs$-twist.
We pick $\zs$ and $\tau$ from the following list of choices:
\begin{enumerate}
\item
$\tau=i$ and $\zs=\zs'\equiv\left(\begin{array}{rr}
 0 & -1 \\ 1 & 0 \\ \end{array}\right)$;
\item
$\tau=e^{\pi i/3}$ and
either
$\zs=\zs''\equiv\left(
\begin{array}{rr} 1 & -1 \\ 1 & 0 \\ \end{array}
\right)\in\SL(2,\Z)$ or
$\zs=-\zs''=\zs''{}^4\equiv\left(
\begin{array}{rr} -1 & 1 \\ -1 & 0 \\ \end{array}
\right)\in\SL(2,\Z)$;
\end{enumerate}
All other possible $\zs$'s are $\SL(2,\Z)$-conjugate
to those in the list, or their inverses (which give
theories that are physically equivalent after a parity
transformation).

Why should we believe that an $\SL(2,\Z)$-twist is allowed?
Duality symmetries have been used quite extensively
in string theory to twist boundary conditions.
(See for instance \cite{Vafa:1996xn}-\cite{Vegh:2008jn}
 for some old and some recent examples.)
Moreover,
The Euclidean partition function on $S^1\times\Mf_3$
with the $\zs$-twist is easy to define --- we simply
treat $S^1$ with radius $R$ as the (Euclidean) time direction 
and calculate $\tr\{(-1)^F e^{-2\pi R\Ham}\hat{\SKer}\}$,
where $F$ is the fermion number, $\Ham$ is the Hamiltonian on
$\Mf_3$, and $\hat{\SKer}$ is the operator that corresponds
to the action of $\zs$ on the Hilbert space.
(Here, we momentarily think of $x_3$ as temporal,
but from now on, until otherwise stated,
we return to thinking about it as a spatial direction.)
Moreover, the $\zs$-twist can be given a purely geometrical
construction in terms of the 5+1D $(2,0)$-theory.
As Witten's conjecture goes \cite{Witten:1995zh},
3+1D \SUSY{4} $U(n)$ SYM with coupling constant $\tau$ can be realized
by compactifying the 5+1D $(2,0)$ theory
(the low-energy limit of $n$ coincident M5-branes
\cite{Strominger:1995ac}) on a $T^2$ with
complex structure $\tau$ and area $\Area$ and taking the 
$\Area\rightarrow0$ limit.
The $T^2$ can be described as the quotient $\C/(\Z+\Z\tau)$
(where $\C$ is the complex plane and $\Z+\Z\tau$ is
the lattice generated by $1$ and $\tau$), and if
$z\sim z+1\sim z+\tau$ is a complex coordinate on $\C$,
then the $\SL(2,\Z)$ duality transformation
$\zs=\begin{pmatrix} \xa & \xb \\
\xc & \xd \\ \end{pmatrix}$
is realized as a change of basis $1\mapsto \xc\tau+\xd$
and $\tau\mapsto \xa\tau+\xb.$ If $\tau$ is fixed by $\zs$
then $\xa\tau+\xb = (\xc\tau+\xd)\tau$ and it is then
easy to check that, assuming $\Imx\tau>0$,
 $|\xc\tau+\xd|=1$, so we can write
\be
\xc\tau+\xd = e^{i\pht}\,,\label{phaseeq}
\ee
for some phase $\pht.$
For $\tau=i$ and $\zs=\zs'$ we have $\pht = \frac{\pi}{2}$;
for $\tau=e^{\pi i/3}$ and $\zs=\zs''$ we have
$\pht= \frac{\pi}{3}$ and for $\zs=-\zs''$ we have
$\pht= \frac{4\pi}{3}.$
The $\zs$-duality transformation can then be realized
as rotation of the $\C$ plane by an angle $\pht$,
which preserves the lattice $\Z+\Z\tau.$

A geometric realization
of the $\zs$-twisted compactification of \SUSY{4} SYM
is now clear.
We take a circle $S^1$ of radius $R$,
parameterized by $0\le x_3<2\pi R$ (we reserve $x_0,x_1,x_2$
for coordinates on $\Mf_3$), and compactify the $(2,0)$
theory on the space parameterized by
$(x_3,z)$ with identifications
\be\label{eqn:Idx3z}
(x_3,z)\sim (x_3,z+1)\sim(x_3,z+\tau)
\sim (x_3+2\pi, e^{i\pht} z),
\ee
and metric
\be\label{eqn:dsx3z}
ds^2 = dx_3^2 + \frac{\Area}{\Imx\tau} |dz|^2\,.
\ee
In the limit $\Area\rightarrow 0$ we recover the \SUSY{4}
$U(n)$ SYM theory compactified on $S^1$ with an $\zs$-twist.

As it stands,
compactification of \SUSY{4} SYM on $S^1$ with an $\zs$-twist
breaks all the supersymmetries.
The supercharges $Q_{a\a}$ ($a=1,\dots, 4$ and $\a=1,2$)
are in the representation
$\rep{2}$ of the Lorentz group $SO(3,1)$ and
$\rep{\overline{4}}$ of the R-symmetry group $SU(4)_R$,
and their complex conjugates
$\bQ^a_\dta$ are in the $\rep{2}'$ of $SO(3,1)$
and $\rep{4}$ of $SU(4)_R.$
As explained in \cite{Kapustin:2006pk},
the supercharges transform under $\zs$ as
\be\label{eqn:zsQ}
\zs: Q_{a\a}\rightarrow
\left(\frac{\xc\tau+\xd}{|\xc\tau+\xd|}\right)^{1/2}Q_{a\a}
 = e^{\frac{i\pht}{2}}Q_{a\a}
\,.
\ee
This is easy to see from the $(2,0)$-theory realization mentioned
above---the duality corresponds to rotation by 
an angle $\pht$ on a plane of the $T^2.$
Since no linear combination of the supercharges is
invariant under $\zs$-duality, no supersymmetry is preserved
by the $\zs$-twist.
We would, however, like to preserve some amount of supersymmetry
so as to be able to use Witten-index techniques later on.
For this purpose we will now add an R-symmetry twist.
As a bonus, we will see that an appropriate twist 
also eliminates some unwanted zero modes of scalar fields.

\subsection{R-symmetry twist}
\label{subsec:Rtwist}
The fields of 3+1D \SUSY{4} $U(n)$ SYM are as follows:
a gauge field $A_\mu$, $6$ adjoint-valued scalar fields
$\Phi^I$ ($I=1,\dots,6$), and $4$ spinor fields
$\psi^a_\a$ ($a=1,\dots,4$ and $\a=1,2$)
and their complex conjugates $\bpsi_{a\dta}$
($a=1,\dots,4$ and $\dta=\dot{1},\dot{2}$).

The R-symmetry is $\Spin(6)=SU(4).$
The  $6$ scalars $\Phi^I$
are in the real representation $\rep{6}$,
the fermions $\psi^a_\a$ are in the complex representation
$\rep{4}$, and $\bpsi_{a\dta}$ are in $\rep{\overline{4}}.$
We pick bases of $\rep{4}$
and $\rep{\overline{4}}$ such that a diagonal element
\be\label{eqn:gtwgen}
\gtw\equiv
\begin{pmatrix}
e^{i\ftw_1} & &  & \\
& e^{i\ftw_2} &  & \\
& & e^{i\ftw_3}  & \\
& & & e^{i\ftw_4}  \\
\end{pmatrix} \in SU(4)_R\,,
\qquad
\left(\sum_a\ftw_a = 0\right)\,,
\ee
acts as
$$
\gtw(\psi^a_\a) = e^{i\ftw_a}\psi^a_\a\,,
\qquad
\gtw(\bpsi_{a\a}) = e^{-i\ftw_a}\bpsi_{a\a}\,.
$$
We also pick a basis of $\rep{6}$ such that
the representation of $\gtw$ in $SO(6)$
acts on the scalars as
\be\label{eqn:gtwPhiPhi}
\left.\begin{array}{rl}
\gtw(\Phi^{2a-1}) &=
\Phi^{2a-1}\cos(\ftw_a+\ftw_4)
-\Phi^{2a}\sin(\ftw_a+\ftw_4)
\\
\gtw(\Phi^{2a}) &=
\Phi^{2a-1}\sin(\ftw_a+\ftw_4)
+\Phi^{2a-1}\cos(\ftw_a+\ftw_4)
\\
\end{array}
\right\}
\ee
for $a=1,2,3$.
On occasion,
we will suppress the indices on $\psi,\bpsi,\Phi$ and denote
the action of $\gtw$ on the fields by $\psi^\gtw,$
$\bpsi^\gtw,$ and $\Phi^\gtw.$

Let $0\le x_3< 2\pi R$ be a periodic coordinate on $S^1.$
We now augment the $\zs$-twist from \secref{subsec:Stwist}
by an R-symmetry twist as follows.
(See also \cite{Witten:1993jg} for a similar use of R-symmetry
in a different context.)
Without the $\zs$-twist, an R-symmetry twist is simply
a modification of the boundary conditions
for the scalars and spinors of \SUSY{4} SYM to
\be\label{eqn:phipsibc}
\Phi(2\pi R) = \Phi(0)^\gtw\,,
\qquad
\psi(2\pi R) = \psi(0)^\gtw\,,
\ee
where $\gtw$ is some element of the R-symmetry group,
and only the $x_3$ argument is shown in \eqref{eqn:phipsibc}.
This twist is independent of the position of the origin $x_3=0$,
and we can combine it with the $\zs$-twist by modifying 
the boundary conditions anywhere along $S^1.$ 

Let us now discuss the amount of supersymmetry that is preserved.
For a generic $\gtw$ in the form \eqref{eqn:gtwgen},
no supersymmetry is preserved.
Some supersymmetry can be preserved
for special choices of the phases $\ftw_a$ ($a=1,\dots,4$)
in \eqref{eqn:gtwgen}.
The combined effect of $\zs$ and $\gtw$ on
a supersymmetry generator $Q_{a\a}$ is given by
\be\label{eqn:Qzsgtw}
Q_{a\a}\rightarrow
e^{\frac{i\pht}{2}-i\ftw_a}Q_{a\a}
\,.
\ee
In general, we get $N=2r$ supersymmetry in 3D, where $r$ is the number
of indices $a$ for which $e^{i\ftw_a}=e^{i\pht/2}$,
according to \eqref{eqn:Qzsgtw}.
We thus can get as high as \SUSY{6} in 3D.
This maximal amount of supersymmetry arises with
\be\label{eqn:gtwN=6}
\gtw=
\begin{pmatrix}
e^{\frac{i}{2}\pht} & &  & \\
& e^{\frac{i}{2}\pht} &  & \\
& & e^{\frac{i}{2}\pht}  & \\
& & & e^{-\frac{3i}{2}\pht}  \\
\end{pmatrix} \in SU(4)_R\,.
\ee
We get \SUSY{4} with
\be\label{eqn:gtwN=4}
\gtw=
\begin{pmatrix}
e^{\frac{i}{2}\pht} & &  & \\
& e^{\frac{i}{2}\pht} &  & \\
& & e^{-i(\pht+\ftw_4)}  & \\
& & &  e^{i\ftw_4} \\
\end{pmatrix} \in SU(4)_R\,,
\ee
for any choice of $\ftw_4$,
and we get \SUSY{2} with
\be\label{eqn:gtwN=2}
\gtw=
\begin{pmatrix}
e^{\frac{i}{2}\pht} & &  & \\
& e^{-i(\ftw_3+\ftw_4-\tfrac{1}{2}\pht)} &  & \\
& & e^{i\ftw_3}  & \\
& & &  e^{i\ftw_4} \\
\end{pmatrix} \in SU(4)_R\,,
\ee
for any choice of $\ftw_3,\ftw_4.$

To summarize, our setting is \SUSY{4} SYM compactified on $S^1$
with an S-duality twist $\zs$ and an R-symmetry twist $\gtw$ along the circle. 
We are interested in the limit
where the radius of the circle $R$ shrinks to zero.
For the rest of this paper, we will take
the twist \eqref{eqn:gtwN=6} which preserves the maximal
amount of \SUSY{6} supersymmetry.

\subsection{Zero modes}
\label{subsec:ZModes}
Our goal is to analyze the low-energy limit of the
compactification of \SUSY{4} $U(n)$ SYM on $S^1$
with both S-duality twist discussed
in \secref{subsec:Stwist} and R-symmetry twist discussed in
\secref{subsec:Rtwist}.
We would like to claim that, for sufficiently small values of $n$,
the low-energy limit leads
to a nontrivial 2+1D topological field theory.
The restriction on $n$ comes because for large values of $n$ the
claim is defeated by the presence of zero modes as we shall
now discuss.

We can attempt to construct a ``Higgs phase'' of the theory
as follows.
If the radius $R$ of $S^1$ is sufficiently large
(compared to a length scale to be defined shortly),
we can first reduce the 3+1D \SUSY{4} theory to its
Coulomb branch, which at a generic point of the moduli space
is described by $n$ free \SUSY{4} $U(1)$ vector multiplets.
We will denote the scalars in the $k$-th multiplet by
$\phi_k^I$ ($k=1,\dots,n$ and $I=1,\dots,6$)
and the electric and magnetic fields in the same multiplet by
$\vE_k$ and $\vB_k,$ respectively.
As in \secref{subsec:Rtwist}, we will suppress
the R-symmetry index $I$ of the scalars.
The next step is to compactify this $U(1)^n$ gauge theory
on $S^1$ with an $\SL(2,\Z)$-duality and R-symmetry twist.
Assuming that
$$
\sum_{I=1}^6|\langle\Phi_k^I\rangle
-\langle\Phi_l^I\rangle|^2\gg\frac{1}{R^2}\,,
\qquad 1\le k<l\le n
$$
(where $\langle\Phi_k^I\rangle$ denote the eigenvalues of the VEV of
$\Phi^I$),
the two-step reduction to low energy is self-consistent.

At first look,
the combined $\SL(2,\Z)$ and R-symmetry twists
set the following low-energy boundary conditions:
\begin{gather*}
\vE_k(2\pi R) = \xa\vE_k(0)+\xb\vB_k(0)\,,
\quad
\vB_k(2\pi R) = \xc\vE_k(0)+\xd\vB_k(0)\,,
\\
\phi_k(2\pi R) = \phi_k(0)^\gtw\,,
\end{gather*}
where
$\zs\equiv\begin{pmatrix} \xa & \xb \\ \xc & \xd \\ \end{pmatrix}$
is the appropriate $\SL(2,\Z)$ element from
\secref{subsec:Stwist} and $\gtw$ is R-symmetry
element from \eqref{eqn:gtwN=6}, which acts on the
suppressed $I$ index of $\phi_k^I.$
The first two boundary conditions can also be written as
$$
\vE_k(0) -\tau\vB_k(0)
=(\xc\tau+\xd)(\vE_k(2\pi R)-\tau\vB_k(2\pi R))
=e^{i\pht}(\vE_k(2\pi R)-\tau\vB_k(2\pi R))\,.
$$

The zero modes can be found by taking the fields
to be independent of the $S^1$ coordinate,
setting $\vE_k(2\pi R)=\vE_k(0)$,
$\vB_k(2\pi R)=\vB_k(0)$, and $\phi_k(2\pi R)=\phi_k(0).$
Since none of the $\SL(2,\Z)$ twists discussed in
\secref{subsec:Stwist} (neither $\zs'$ nor $\pm\zs''$)
have an eigenvalue $1$, none of the phases $\pht$ of \eqref{phaseeq}
are zero, and there are no zero modes of
the vector fields with these boundary conditions.
Since $\gtw$ of \eqref{eqn:gtwN=6}
has no eigenvalue $1$, there are also no zero modes
of the scalar fields.

However, when the gauge symmetry is broken as
$U(n)\rightarrow U(1)^n$ by the VEVs $\langle\Phi_k^I\rangle$,
we also get an action of the Weyl group $S_n\subset U(n)$
on the low-energy fields. It acts by permuting the indices
$k=1,\dots,n.$ Since $S_n$ is a remnant of the gauge group $U(n)$,
we are allowed to consider sectors for which the boundary
conditions along $S^1$ are twisted by an element $\sigma\in S_n.$
The boundary conditions in this sector become
$$
\vE_{\sigma(k)}(0) -\tau\vB_{\sigma(k)}(0)
=e^{i\pht}(\vE_k(2\pi R)-\tau\vB_k(2\pi R))\,,
\qquad
\phi_{\sigma(k)}(0)^\gtw = \phi_k(2\pi R)\,,
$$
where $\sigma$ is understood as a permutation map
$\sigma:\{1,\dots,n\}\rightarrow\{1,\dots,n\}.$
We get a surviving 2+1D low-energy mode of the vectors
for any linear combination
$\sum_{k=1}^n C_k (\vE_k -\tau\vB_k)$ such that the coefficients
satisfy
\be\label{eqn:Cksigma}
C_k = e^{i\pht}C_{\sigma(k)}\,,\qquad k=1,\dots,n.
\ee
In other words, $e^{i\pht}$ must be an eigenvalue of $\sigma$
in its fundamental representation.

Similarly, from \eqref{eqn:gtwN=6} we see that the
eigenvalues of $\gtw$ in the representation $\rep{6}$
of $SO(6)$ are $e^{\pm i\pht}$ (each one occurring with
a multiplicity of $3$). Thus, we also get three complex scalar
zero modes
of the form $\sum_{I=1}^6\sum_{k=1}^n \lambda^I_{(a)} C_k\phi^I_k.$
Here $\lambda^I_{(a)}$, for $a=1,2,3$, are the three  eigenvectors
of $\gtw$ with eigenvalue $e^{-i\pht}$ (which cancels the phase of 
\eqref{eqn:Cksigma} to give rise to the zero modes).
In the basis of \eqref{eqn:gtwPhiPhi}, we can take
$$
\sum\lambda^I_{(1)}\phi^I = \phi^1 + i \phi^2\,,\qquad
\sum\lambda^I_{(2)}\phi^I = \phi^3 + i \phi^4\,,\qquad
\sum\lambda^I_{(3)}\phi^I = \phi^5 + i \phi^6\,.
$$
for any set of coefficients
$\{C_k\}$ that satisfy \eqref{eqn:Cksigma}.

For $\tau=i$ and $\zs=\zs'$ we have $\pht = \frac{\pi}{2}$,
and a nonzero solution to \eqref{eqn:Cksigma} exists
only if the Weyl group has an element of order four, and hence only if $n\ge 4.$
For the threshold value $n=4$, a nonzero solution requires that
$\sigma$ act as $\sigma:(1,2,3,4)\mapsto(2,3,4,1)$, up to conjugation. 
The space of solutions is then one-dimensional,
and the low-energy limit of the compactified \SUSY{4} theory
is given by a single 2+1D free
\SUSY{8} multiplet of $8$ real scalar fields
(the three complex scalar zero modes
give $6$ fields, and the vector field gives rise to
$2$ scalar fields)
with moduli space $(\C^3\times T^2)/\Z_4.$
The $\C^3$ factor corresponds to the three scalar zero modes,
and the $T^2$ (with complex structure $\tau=i$)
factor comes out of the vector zero mode.
The action of $\Z_4$ corresponds to multiplication
by $i$ for each of the four factors in
$\C\times\C\times\C\times T^2$, and is derived from the
identification
$$
\sum_I\sum_k \lambda^I_{(a)}C_k\phi^I_k
\sim\sum_I\sum_k \lambda^I_{(a)}C_k\phi^I_{\sigma(k)}
=\sum_I\sum_k \lambda^I_{(a)}C_{\sigma^{-1}(k)}\phi^I_{k}
=e^{i\pht}
\sum_I\sum_k \lambda^I_{(a)}C_{k}\phi^I_{k}\,.
$$

For $\tau=e^{\pi i/3}$ and $\zs=\zs''$ we have
$\pht= \frac{\pi}{3}$ and therefore no zero modes
for $n<6$, while for for $\zs=-\zs''$ we have
$\pht= \frac{4\pi}{3}$ and there are no zero modes if $n<3.$
The analysis of the threshold cases is similar to the one above.

Thus, when we reduce the \SUSY{4} theory to its low-energy
limit and then compactify with a twist, we find no
low-energy zero-modes for
\begin{itemize}
\item
$\tau=i$ and $\zs=\zs'$ if $n=1,2,3$;
\item
$\tau=e^{\pi i/3}$ and $\zs=\zs''$ if $n=1,2,3,4,5$;
\item
$\tau=e^{\pi i/3}$ and $\zs=-\zs''$ if $n=1,2$.
\end{itemize}

We can also provide an alternative argument that does not
use the low-energy limit of \SUSY{4} SYM.
We can glean more information
about the putative Coulomb branch of the low-energy
2+1D theory by studying the VEV of BPS operators.
Set $\Zphi\equiv\phi^5 + i \phi^6$ and consider
the BPS operators
$$
\Op_p\equiv\gYM^{-p}\tr(\Zphi^p)\,,\qquad p=1,2,\dots
$$
According to \cite{Intriligator:1998ig},
with this normalization the operators are
$\SL(2,\Z)$-duality invariant.
The action of $\gtw$ is
$$
(\Op_p)^\gtw = e^{i p\pht}\Op_p\,.
$$
It follows that $\Op_p$ is single-valued
in our setting if and only if $e^{i p \pht}=1.$
Therefore,
\begin{itemize}
\item
for $\tau=i$ and $\zs=\zs'$, $\langle\Op_p\rangle\neq 0$
requires $p\in 4\Z$;
\item
for $\tau=e^{\pi i/3}$ and $\zs=\zs''$
$\langle\Op_p\rangle\neq 0$ requires $p\in 6\Z$;

\item
for $\tau=e^{\pi i/3}$ and $\zs=-\zs''$
$\langle\Op_p\rangle\neq 0$ requires $p\in 3\Z$.
\end{itemize}
For $U(n)$, $\Op_{n+1},\Op_{n+2},\ldots$ are not independent
of $\Op_1,\dots,\Op_n.$ Thus for $\tau=i$ and $\zs=\zs'$, for example,
if $n<4$ none of the operators $\Op_p$ can get a VEV.
By studying additional BPS operators we can restrict
the form of the Coulomb branch and reach the same conclusion
as above that for $n<4$ there are no zero modes.

In the rest of this paper, we take the gauge group to have low enough
rank so that after S-duality and R-symmetry twists there are no zero modes
in the compactified theory.

\section{S-duality kernel and topological field theory}
\label{sec:Kernel}
In the limit $R\rightarrow 0$, the construction
in \secref{sec:problem} essentially reduces to
a question about the duality transformation defined
by the $\SL(2,\Z)$ element $\zs.$
In order to see what this question is more precisely,
it is convenient to think about the $S^1$ as a (Euclidean)
time direction, and think about the $\zs$-twist as an
insertion of an operator that realizes the duality.
To explore this point of view further,
we now switch to a formal Schr\"odinger representation.

\subsection{Schr\"odinger representation}
\label{subsec:Schroedinger}
We will assume that the theory is formulated on a
compact three-manifold $\Mf_3$, so that the full spacetime is
$\R\times\Mf_3.$
We will use the convention that $i,j,\dots=1,2,3$ are spatial indices
and $\mu,\nu,\dots=0,1,2,3$ are spacetime indices.
We will denote the 3D metric on $\Mf_3$ by $g_{ij}.$
The full metric will be $ds^2=-dt^2 + g_{ij}dx^i dx^j.$
In this subsection, we will not restrict $\gYM$ and
$\theta$, and work with a generic coupling constant
$$
\tau = \frac{4\pi i}{\gYM^2} +\frac{\theta}{2\pi}\ .
$$

We will work in the Hamiltonian formalism
and in the temporal gauge
$$
A_0=0.
$$
The spatial components of the gauge field will be denoted by
the 1-form $A\equiv A_i dx^i$ that is to be understood as defined on $\Mf_3$ at a fixed time.
Thus, by $dA$ we will always mean ``the
exterior derivative on $\Mf_3$''
so that $dA$ is a 2-form on $\Mf_3$. In addition to the gauge field,
we have 6 adjoint-valued scalar fields
$\Phi^I$ ($I=1,\dots,6$) and 4 spinor fields
$\psi^a_\a$ ($a=1,\dots,4$ and $\a=1,2$)
and their complex conjugates $\psi_{a\dta}$
($a=1,\dots,4$ and $\dta=\dot{1},\dot{2}$).
We will denote the collective configuration field by
$\lV\equiv\{A,\psi^a_\a,\Phi^I\}.$
Physical states are then formally represented by gauge invariant
wavefunctions $\wvF(\lV)\equiv\wvF\{A,\psi^a_\a,\Phi^I\}.$

We denote the ($\frak{u}(n)$ or $\frak{su}(n)$ Lie algebra valued)
vector field canonically dual to $A$ by $E^i\px{i}.$
We understand it as the operator
$$
E^i\equiv -2\pi i\frac{\delta}{\delta A_i}\ 
$$
acting on wavefunctions.
The magnetic field is $B_i dx^i = *dA$, where $*$ is the three-dimensional Hodge star operator.
The Hamiltonian is
\be\label{eqn:Ham}
\Ham = \int \sqrt{g} d^3x\tr\{
\frac{1}{2}\gYM^2 g_{ij}E^i E^j
+\frac{1}{2\gYM^2} g^{ij} B_i B_j
+\cdots
\}\ .
\ee

The conjectured S-dual description involves dual fields:
$\sdA_i$, $\sdpsi^a_\a$, $\sdPhi^I$, which will be 
collectively denoted by $\sdlV$.
The dual coupling constant and $\theta$-angle are given by
$$
\frac{4\pi i}{\sdgYM^2} +\frac{\sdtheta}{2\pi}\equiv \sdtau
=\frac{\xa\tau+\xb}{\xc\tau+\xd}\ .
$$
The dual Hamiltonian will be denoted by $\sdHam.$

\subsection{S-duality kernel}
\label{subsec:Sker}
Formally, the Hamiltonian \eqref{eqn:Ham} acts
on the Hilbert space of gauge-invariant wavefunctions.
We are interested in how the $\SL(2,\Z)$ group of dualities
are realized in the Hamiltonian formalism.

The group $\SL(2,\Z)$ is generated by
$\begin{pmatrix} 0 & -1 \\ 1 & 0 \\ \end{pmatrix}$
and
$\begin{pmatrix} 1 & 1 \\ 0 & 1 \\ \end{pmatrix}.$
The latter corresponds to a shift $\theta\rightarrow\theta+2\pi$
and  acts on the wavefunction in a simple way:
\be\label{eqn:CSshift}
\wvF(\lV)\rightarrow e^{i I_{CS}(A)}\wvF(\lV)\,,
\ee
where
$$
I_{CS}(A)\equiv
\frac{1}{4\pi}\int\tr\{A\wedge dA + \frac{2}{3}A\wedge A\wedge A\}d^3x
$$
is the level-$1$ Chern--Simons action.
Equations \eqref{eqn:CSshift} can be seen
either by directly integrating  the extra $F\wedge F$
term in the action that results from the shift
$\theta\rightarrow\theta+2\pi$, or by checking that
it acts on the electric field operator in the appropriate way:
$E_i\rightarrow E_i + B_i.$

S-duality acts on the wavefunction in a more complicated
and generally unknown way.
We will denote  the S-duality operator acting on the Hilbert space by $\OpS.$
We have
\be\label{eqn:SH}
\OpS\Ham = \sdHam\OpS\,.
\ee
Formally, acting on $\wvF(\lV)$,
it produces a dual wavefunction
$\sdwvF(\sdlV)\equiv\sdwvF\{\sdA,\sdpsi^a_\a,\sdPhi^I\}.$
We can represent S-duality
by a (Fredholm) kernel $\SKer(\lV,\sdlV)$
that acts as
$$
\sdwvF(\sdlV) = \int \SKer(\lV,\sdlV)\wvF(\lV)[\cD\lV],
$$
where $[\cD\lV]$ denotes a path integral on all the fields
$A,\psi^a_\a,\Phi^I$, and $\SKer$ is a (generally nonlocal)
functional of both field configurations, $\lV$ and $\sdlV$,
which is separately invariant under
a gauge transformation of $\lV$ and a gauge transformation
of $\sdlV$.\footnote{The wavefunction $\wvF$
is closely related, after Wick rotation,
to the boundary conditions defined in \cite{Gaiotto:2008sa}.
For example, Dirichlet boundary conditions on the gauge fields
correspond to a wavefunction $\delta(B_i)$, and
Neumann boundary conditions correspond to a constant wavefunction
(on which $\delta/\delta A_i\rightarrow 0$). More complicated
wavefunctions were constructed in \cite{Gaiotto:2008sa} 
by coupling the gauge degrees of freedom to extra boundary (3D)
degrees of freedom. Our $\SKer$ is directly related to
the action of S-duality on these boundary conditions
as described in \cite{Gaiotto:2008ak}.
}

Let us now discuss the possible ambiguities in our definition
of $\SKer.$
The S-duality operator $\OpS$ transforms eigenstates
of the Hamiltonian $\Ham$ to eigenstates of $\sdHam$,
but we can consider multiplication by unitary operators
that commute with the Hamiltonians.
In other words, let $\xP$ be some unitary operator that commutes with $\Ham$
and let $\sdxP$ be another unitary operator that commutes
with $\sdHam.$
We are considering the freedom to change
$\OpS\mapsto\sdxP\OpS\xP.$
In the following discussion we will argue that there are
essentially no nontrivial ambiguities except
for a global phase and time translation.
We will use the fact that $\OpS$ commutes with the Hamiltonian
(in the sense of \eqref{eqn:SH})
and with R-charge, and takes local operators of \SUSY{4} SYM
to local operators of the dual theory.

Since the unitary operator $\xP$ commutes with the Hamiltonian $\Ham$,
it must be a function of the conserved charges
which are $\Ham$ and the R-charge
operators, and similarly $\sdxP$ must be a function
of $\sdHam$ and R-charge of dual theory. Since $\OpS$ commutes
with the $15$ R-charge generators, $\xP$ and $\sdxP$
can only depend on $SU(4)$-invariant combinations
of the $SU(4)$-generators, i.e., on the Casimirs of $SU(4).$
For a generic  positively curved
metric $g_{ij}$ and generic $SU(4)_R$ bundle
over a compact $\Mf_3$, we expect the energy levels
to fall into $SU(4)_R$ multiplets, but other than that
to be discrete and nondegenerate.
Thus, for a fixed metric and R-bundle,
we may assume that $\xP$ and $\sdxP$ only depend
on $\Ham$ and $\sdHam$, respectively, but not on the R-charge.
(The Casimirs of the $SU(4)$ representation
can be absorbed in a function of the energy,
as the representation can generically be read off from the energy.)
We can then set $\xP=f(\Ham)$ and $\sdxP=g(\sdHam)$
for some functions $f,g$ of the discrete energy eigenvalues.

We now have $\sdxP\OpS\xP = \OpS g(\Ham)f(\Ham)$,
and we can assume without loss of generality that $\sdxP=1.$
We also know that $\OpS$ takes local operators $\Op$
to local operators $\OpS\Op\OpS^{-1}$.
To preserve this propetry,
$\OpS\xP\Op\xP^{-1}\OpS^{-1}$ must also be local.
Here, we are considering locality in time as well as
in space. This restricts $\xP$ to be of
the form $e^{i T\Ham+i\phi}$ for some constant $T$ and
constant phase $\phi.$ 
This reflects the obvious fact that there is no canonical
way to identify the original time and the time in the dual theory,
and so the S-duality has an undetermined time-translation in it.

We can do better when the coupling constant $\tau$ takes 
one of the S-dual values discussed in \secref{subsec:Stwist}
and $\hat{\SKer}$ realizes the action of corresponding element $\zs\in\SL(2,\Z)$.
Then, we can identify the original Hilbert space
with the dual Hilbert space,
and also require that the ground state transforms into itself,
without a phase. This completely eliminates all the ambiguity.

\subsection{Metric independence}
\label{subsec:Metric}
We will now argue that $\SKer$ is topological.
To show that, we need to check that $\SKer$
is independent of the metric $g_{ij}.$
Consider a small deformation
$\delta g_{ij}$ of the metric on $\Mf_3.$
The corresponding corrections to $\Ham$ and $\sdHam$ are
$$
\delta\Ham = \int_{\Mf_3}\sqrt{g}\delta g_{ij} T^{ij} d^3x
\,,\qquad
\delta\sdHam = \int_{\Mf_3}\sqrt{g}\delta g_{ij} \sdT^{ij} d^3x
\,,
$$
where $T^{ij}$
are the spatial components of the energy-momentum tensor
of the original \SUSY{4} theory, and $\sdT^{ij}$
are the components of the energy-momentum tensor
of the dual theory.
Since S-duality maps the energy-momentum tensor to itself,
$$
\OpS(\delta\Ham) = (\delta\sdHam)\OpS
\,.
$$

Let $\delta\OpS$ be the small
change in $\OpS$ due to the metric deformation above.
Then,
$$
(\OpS+\delta\OpS)(\Ham+\delta\Ham)
=(\sdHam+\delta\sdHam)(\OpS+\delta\OpS)
\Longrightarrow
(\delta\OpS)\Ham-\sdHam(\delta\OpS) = 0
\,.
$$
Thus, $\delta\OpS$ transforms an eigenstate of $\Ham$
to an eigenstate of $\sdHam$ with the same energy eigenvalue.
Then $\OpS^{-1}(\OpS+\delta\OpS)$ commutes with the
Hamiltonian. We can similarly argue
that it commutes with the R-charge.
In addition, if $\Op$ is a local operator in the original
Hilbert space,
then so is $\OpS^{-1}(\OpS+\delta\OpS)\Op(\OpS+\delta\OpS)^{-1}\OpS.$
By the same argument as above it then follows that
$\OpS^{-1}(\OpS+\delta\OpS)$ is the identity operator
(up to a possible time translation and a global phase), which proves the claim.
A similar argument shows that $\SKer$ is also independent
of the coupling constant $\tau.$  To show that, we can use the fact
that the dilaton operator $\partial\Ham/\partial\tau$
maps to itself under S-duality.

The construction in \secref{sec:problem} is related
to the S-kernel $\SKer$ in the following way.
Take one of the selfdual values of $\tau$ and the corresponding
element $\zs\in\SL(2,\Z)$ as in \secref{subsec:Stwist}, and
let the S-duality and R-symmetry twists be at $x_3=0.$
Suppose we insert some Wilson-line operator
$\Op$ at $x_3=\epsilon>0$ with $\epsilon\ll R.$
Its expectation value in
the theory of \secref{sec:problem} is given by
$\tr\{(-1)^F e^{-2\pi R\Ham}\Op\OpS\gtw\}$, 
where $\hat{\SKer}$ represents
the action of $\zs$ on the Hilbert space. 
If we could take the naive limit $R\rightarrow 0$,
we would get $\tr\{(-1)^F \Op\OpS\gtw\}$ which corresponds to
calculating the expectation value of $\Op$ in
a 3D theory whose action is formally given by
\be\label{eqn:AcV}
\AcV(\lV)\equiv -i\log\SKer(\lV,\lV^\gtw)\,,
\ee
where  $\lV^\gtw$ are the fields $\lV$
after the R-symmetry twist $\gtw.$
In other words,
the expectation value of an operator $F(\lV)$ (say a Wilson
line) in that theory is by definition given by
$$
\langle F\rangle\equiv
\int e^{i\AcV(\lV)}F(\lV)[\cD\lV]
\equiv
\int \SKer(\lV,\lV^\gtw)F(\lV)[\cD\lV]
\,.
$$

However, since $\Ham$ is not bounded from above, taking
the limit $R\rightarrow 0$ is potentially dangerous.
One potential problem is that $\tr\{(-1)^F\Op\OpS\gtw\}$ receives
large contributions from high-energy modes,
for example if $\Op$ is a Wilson line with a cusp.
Another problem is if $\tr\{(-1)^F\Op\OpS\gtw\}$ is ill-defined
because of massless zero-modes. This is the case
if $\gtw$ is not generic enough.
For the time being, we will assume that none of these
problems arise and that the $R\rightarrow 0$ limit is safe.
We will proceed to discuss the diagonal of the S-duality kernel,
as defined in \eqref{eqn:AcV}.

In general, the kernel $\SKer$ depends on two independent field configurations
$\lV$ and $\sdlV$ on $\Mf_3$, and we do not expect it
to be expressible in terms of an integral of a local expression
in the fields. However, the construction in
\secref{sec:problem} makes it clear that
$\AcV(\lV)$ is the action of a local theory.
We further conjecture that $\AcV(\lV)$
is an integral of a local expression in the fields $\lV$,
i.e., the theory that it defines is not only local,
but local in the variables $\lV.$
Since we argued that $\SKer(\lV,\sdlV)$ is topological,
i.e., independent of the metric $g_{ij}$,
we expect $\AcV(\lV)$ to also be topological. We conclude that the diagonal
of the S-duality kernel defines a local topological field theory
in three dimensions.

\subsection{Expectation value of a Wilson-loop pair}
An interesting aspect of the nonlocal structure
$\SKer(\lV,\sdlV)$ is that it renders the correlation
functions of Wilson loops easy to calculate.
For this purpose, we specialize to the $\tau=i$ case, and let the
nonlocal kernel $\SKer(\lV,\sdlV)$ represent the action of $\zs=\zs'\in\SL(2,\Z)$.
For clarity of discussion, we momentarily suppress
the dependence on superpartners and write the kernel
as $\SKer(A,\sdA).$

Choose two loops $\LoopC$ and $\sdLoopC$ in a three-manifold
and define the double-Wilson-loop expectation value:
$$
W(\LoopC,\sdLoopC)=
\int\SKer(A,\sdA)
\tr\left(P e^{i\oint_\LoopC A}\right)
\tr\left(P e^{i\oint_\sdLoopC \sdA}\right)
[\cD A][\cD\sdA]\,,
$$
where $\tr$ is in the fundamental representation of $SU(n).$
We will now present a heuristic calculation of
$W(\LoopC,\sdLoopC).$
We work in the formal Schr\"odinger representation discussed
in \secref{subsec:Schroedinger} and define the operators
$$
\widehat{W}(\LoopC) = \tr\left(P e^{i\oint_\LoopC A}\right)
\,,\qquad
\widehat{W}(\sdLoopC) = \tr\left(P e^{i\oint_\sdLoopC A}\right)
\,.
$$
We also define the formal state $\ket{1}$
which has a formal wavefunctional $\wvF\{A\}=1$ for every
gauge field configuration $A$ (which is related to Neumann
boundary conditions in \cite{Gaiotto:2008sa}).
Then, formally,
$$
W(\LoopC,\sdLoopC) =
\bra{1}\widehat{W}(\sdLoopC)\SKer\widehat{W}(\LoopC)\ket{1}
\,.
$$
Let $\widehat{M}(\sdLoopC)$ be the 't Hooft loop
operator \cite{'tHooft:1977hy} associated with the loop $\sdLoopC.$
Then $\widehat{W}(\sdLoopC)\SKer=\SKer\widehat{M}(\sdLoopC)$
(see \cite{Kapustin:2005py}\cite{Kapustin:2006pk}).
Using the commutation relation \cite{'tHooft:1977hy}
$$
\widehat{M}(\sdLoopC)\widehat{W}(\LoopC)
=\widehat{W}(\LoopC)\widehat{M}(\sdLoopC)
e^{\frac{2\pi i}{n}\Linking(\LoopC,\sdLoopC)}\,,
$$
where $\Linking(\LoopC,\sdLoopC)$ is the {\it linking number}
of the loops $\LoopC$ and $\sdLoopC$,
and using the fact that $\widehat{M}(\sdLoopC)$
acts by changing one gauge configuration to another,
so that $\widehat{M}(\sdLoopC)\ket{1}=\ket{1}$, we get
\bear
W(\LoopC,\sdLoopC) &=&
\bra{1}\widehat{W}(\sdLoopC)\SKer\widehat{W}(\LoopC)\ket{1}
=\bra{1}\SKer\widehat{M}(\sdLoopC)\widehat{W}(\LoopC)\ket{1}
=e^{\frac{2\pi i}{n}\Linking(\LoopC,\sdLoopC)}
\bra{1}\SKer\widehat{W}(\LoopC)\widehat{M}(\sdLoopC)\ket{1}
\nn\\ &=&
e^{\frac{2\pi i}{n}\Linking(\LoopC,\sdLoopC)}
\bra{1}\SKer\widehat{W}(\LoopC)\ket{1}
=e^{\frac{2\pi i}{n}\Linking(\LoopC,\sdLoopC)}
\bra{1}\widehat{M}(\LoopC)\SKer\ket{1}
=e^{\frac{2\pi i}{n}\Linking(\LoopC,\sdLoopC)}
\bra{1}\SKer\ket{1}
\,,
\nn
\eear
The last equality is justified by inserting $\int[\cD A]\ket{A}\bra{A}$ 
in front of $\SKer$ on both sides and using 
$\bra{1}\widehat{M}(\LoopC)\ket{A}=1=\langle 1 | A \rangle$
for any configuration eigenstate $\ket{A}$, according to the definition of 
$\ket{1}$. 

The normalization factor $\bra{1}\SKer\ket{1}$
is independent of the loops $\LoopC,\sdLoopC$,
and assuming that it can be regularized to a nonzero value,
we find a topological result:
\be\label{eqn:WCC}
W(\LoopC,\sdLoopC)\propto
e^{\frac{2\pi i}{n}\Linking(\LoopC,\sdLoopC)}\,.
\ee
In principle, this result can be used to reconstruct
the nonlocal kernel $\SKer(A,\sdA)$,
at least on a lattice. It would be interesting
to see if this can lead to a useful expression for $\SKer$
\cite{InProgress}.

\subsection{Electric and magnetic fluxes}
\label{subsec:EMfluxes}
We close this section by studying how the S-duality operator $\hat{\SKer}$ 
acts on the electric and magnetic fluxes \cite{'tHooft:1977hy}.
Our analysis is based on a review contained in \cite{Kapustin:2006pk},
to which we refer the reader for more information.

Let us start with the gauge group $SU(n)$ and its adjoint form $SU(n)/\Z_n$. 
Since we consider the theory on a manifold
$\MXf=S^1\times\Mf_3$, where we view the $S^1$ as Euclidean time, 
the electric and magnetic fluxes $\ef$ and $\mf$ take values in the following abelian groups:
\be
\ef\in \mathrm{Hom}(H^1(\Mf_3,\Z_n),U(1)),\qquad
\mf\in H^2(\Mf_3,\Z_n)\,.
\ee
As $\mathrm{Hom}(H^1(\Mf_3;\Z_n),U(1))$ is naturally isomorphic to $H^2(\Mf_3;\Z_n)$, 
one can meaningfully talk about exchanging the electric and magnetic fluxes. 
More precisely, the S-duality conjecture states that the fluxes transform as
\be\label{fluxtransf}
\left(\begin{array}{c} \ef \\ \mf \end{array}\right)\rightarrow 
\left(\begin{array}{c} \tilde{\ef} \\ \tilde{\mf} \end{array}\right)
=\begin{pmatrix} \xa & \xb \\ \xc & \xd \\ \end{pmatrix} 
\left(\begin{array}{c} \ef \\ \mf \end{array}\right)
\ee
under the $\SL(2,\Z)$ action.

Each choice of $\ef$ and $\mf$ defines a Hilbert space $\cal{H}_{\ef,\mf}$. 
Its elements are those wavefunctions defined on the space of $SU(n)/\Z_n$ bundles of 
topological type defined by $\mf$ that transform in a way specified by $\ef$ 
under a large gauge transformation. 
In view of \eqref{fluxtransf}, the S-duality operator $\hat{\SKer}$ acts on these Hilbert spaces as
\be
\hat{\SKer}:\cal{H}_{\ef,\mf}\rightarrow\cal{H}_{\tilde{\ef},\tilde{\mf}}\,.
\ee
Therefore, with the choices of gauge coupling $\tau$ and $\zs\in\SL(2,\Z)$ 
that we have considered so far, $\cal{H}_{\ef,\mf}$ is invariant under duality 
in the following cases\footnote{Note that $\ef,\mf$ are elements of abelian torsion groups
where $n\cdot\mf=0$ and $n\cdot\ef=0$ doesn't imply
$\mf=0$ and $\ef=0$.}:
\begin{itemize}
\item for $\tau=i$ and $\zs=\zs'$: $\ef=\mf$ and $2\mf=0$;
\item for $\tau=e^{\pi i/3}$ and $\zs=\zs''$: $\ef=\mf=0$;
\item for $\tau=e^{\pi i/3}$ and $\zs=-\zs''$:
$\ef+\mf=0$ and $3\ef=0$.
\end{itemize}
For example, if $n=2$, then $2\mf=0$ for any $\mf$, so at $\tau=i$, 
the Hilbert space $\cal{H}_{\ef,\ef}$ is invariant for any $\ef$ under the action of $\zs=\zs'$.

In the $U(1)$ theory, the electric and magnetic fluxes take values in $H^2(\Mf_3,\Z)$, 
and also transform as \eqref{fluxtransf} under the duality. 
It is easy to see that the only invariant Hilbert space $\cal{H}_{\ef,\mf}$ 
in this case is the one with $\ef=\mf=0$.

By combining the results for the $SU(n)$ and $U(1)$ cases above, 
we can answer the same question for the $U(n)$ theory. 
We first note that the Hilbert space for the $U(n)$ theory decomposes as
\be
\mathcal{H}^{U(n)}=\bigoplus_{\substack{\ef'=\ef\mod{n} \\
\mf'+\mf=0\mod{n}}}\mathcal{H}^{U(1)}_{\ef',\mf'}
\otimes\mathcal{H}^{SU(n)}_{\ef,\mf}\,.
\ee
The fluxes are correlated because of the $\Z_n$ action in
$U(n)=[SU(n)\times U(1)]/\Z_n$. It follows from the above results that, 
for all values of $\tau$ and $\zs$ under consideration, 
the only invariant Hilbert space $\mathcal{H}^{U(1)}_{\ef',\mf'}
\otimes\mathcal{H}^{SU(n)}_{\ef,\mf}$ is the one with
$(\ef',\mf')=(0,0)$ and $(\ef,\mf)=(0,0)$.

Now, let us return to $SU(n).$
We have seen above that only a small subset of the
possible flux combinations $\ef,\mf$ are allowed.
How can we modify the $\SL(2,\Z)$-twist construction to
include fluxes that are not $\SL(2,\Z)$ invariant?
Suppose the $\SL(2,\Z)$-twist is at $x_3=0$
and that for $x_3<0$ we have fluxes $\ef,\mf$, so that for
$x_3>0$ we have fluxes $\tilde{\ef},\tilde{\mf}$, as in \eqref{fluxtransf}.
To make this construction consistent we need to insert
an operator at some other $x_3$, say $x_3=\epsilon$,
that augments the electric flux by $\ef-\tilde{\ef}$
and augments the magnetic flux by $\mf-\tilde{\mf}.$
Let $\LoopC$ be a loop (or union of loops) in $\Mf_3$
whose homology class is equivalent to
the cohomology class $\ef-\tilde{\ef}.$
Then, an appropriate Wilson loop operator $W(\LoopC)$ inserted
at $x_3=\epsilon$ is the operator we need.
Similarly, a 't Hooft loop for $\LoopC$ that is
Poincar\'e dual to the cohomology class $\mf-\tilde{\mf}$
will augment the magnetic flux by the desired amount.

For our application we especially need to consider
the case $\Mf_3=\RSC_h\times\R$, where $\RSC_h$
is some Riemann surface (of genus $h$).
Let $\ef_0,\mf_0$ be the electric and magnetic fluxes
through $\RSC_h$, which take values in $\Z_n.$
Choose a point $p\in\RSC_h$ and
a representation $\rep{r}$ of $SU(n)$
and consider the Wilson-loop operator
$$
W(\rep{r},p,\epsilon)=\tr_{\rep{r}}
\left(P \exp\oint_{\{p\}\times\R\times\{x_3=\epsilon\}} A\right).
$$
Here, $\{p\}\times\R\times\{x_3=\epsilon\}\subset
\RSC_h\times\R\times S^1$ is the line at $p$ and $x_3=\epsilon.$
This operator $W(\rep{r},p,\epsilon)$ augments
the electric flux by the number of boxes $b$ of the Young diagram
associated with $\rep{r}.$
Thus, we require $b\equiv(\ef_0-\tilde{\ef}_0) \mod n$,
where $\tilde{\ef}_0$ is the  electric flux through
$\RSC_h$ after the $\SL(2,\Z)$-duality $\zs.$
Similarly, a 't Hooft loop at $p$ will suffice if
$b\equiv(\mf_0-\tilde{\mf}_0) \mod n.$

We do not know if there are any further restrictions required
of $\rep{r}$ in either the electric or magnetic case.
We therefore believe, for example, that the compactification
$\RSC_h\times\R\times S^1$
with the $\zs'$ twist at $x_3=0$ in the $S^1$ direction,
with $\mf_0=\ef_0=m$ for $0<x_3<\epsilon$ (and some $m\in\Z_n$)
and with $\mf_0=-\ef_0=m$ for $\epsilon<x_3<2\pi$,
and with a straight
Wilson line at $x_3=\epsilon$ and $p\in\RSC_h$ in a representation
$\rep{r}$ with $2m$ boxes, is consistent.

\section{T-duality twist and geometric quantization}
\label{sec:Ttwist}

Before we approach the main case of interest,
\SUSY{4} SYM with an S-duality twist,
it is useful to study a simpler problem where similar
ideas arise.
The problem that we will study
in this section is the compactification of a 1+1D theory
on $S^1$ with the insertion of a T-duality twist,
assuming the theory is selfdual.
Arguments along the line presented in \secref{sec:Kernel}
suggest that this construction yields a topological
theory in 0+1D.
A 0+1D topological theory is a quantum mechanical system
with a Hamiltonian that is identically zero.
In the examples that we study below, it will have a finite-dimensional
Hilbert space (of vacua).

We will begin with a free scalar at the self-dual radius.
In this case, we will see that the resulting topological theory
is trivial. We will then proceed to a free $\sigma$-model
with target space $T^d$ at a point
in the moduli space that is invariant under some duality
transformation in $O(d,d,\Z).$ We will demonstrate
that the resulting topological theory is equivalent
to geometric quantization of the target space.
We will then comment on a more general case where
we twist a selfdual supersymmetric $\sigma$-model
by the duality that is mirror symmetry.

\subsection{Warm-up: free self-dual scalar}
\label{subsec:freesc}
Consider a free real 1+1D boson $\Phi(\wssig,\wstau)$
where $0\le \wssig<2\pi$ is the spatial coordinate and
$\wstau$ is time.
The action is
$$
\ActS = \frac{1}{4\pi}\int\left\{
(\px{\wstau}\Phi)^2-(\px{\wssig}\Phi)^2
\right\}d\wssig d\wstau
\,,
$$
and the Hamiltonian is
$$
\Ham = \frac{1}{4\pi}\int\left\{
(\px{\wstau}\Phi)^2+(\px{\wssig}\Phi)^2
\right\}d\wssig\,.
$$
We take the boson at the selfdual radius,
so that $\Phi\sim\Phi +2\pi.$

There are many ways to prove T-duality of this simple
free theory \cite{Polchinski}, but for our purposes
we need to do it in the Schr\"odinger representation.
We therefore expand, at fixed $\tau$,
$$
\Phi(\wssig) = w\wssig
+\sum_{n=-\infty}^\infty\phi_n e^{i n\wssig}
\,,
$$
where $w\in\Z$ is the winding number, $\phi_n^*=\phi_{-n}$
are the Fourier modes, and $\phi_0$ is real and periodic with
period $2\pi.$
A state in the Hilbert space is described by a wavefunction, which is a formal expression
$\Psi(w,\{\phi_n\})$, and T-duality acts as
$$
\Psi(w,\{\phi_n\})\rightarrow
\tdlPsi(\tdlw,\{\tdlphi_n\})
=\sum_w\int\prod_n d\phi_n
 \TKer(\tdlw,\{\tdlphi_n\}; w,\{\phi_n\})\Psi(w,\{\phi_n\})
\,,
$$
where the duality kernel $\TKer$ is given by \cite{Lozano:1995aq}:
\bear
\TKer(\tdlw,\{\tdlphi_n\}; w,\{\phi_n\})
&=&\exp\{
i (\tdlw\phi_0-w\tdlphi_0)
+\sum_{n=-\infty}^\infty n \phi_n\tdlphi_{-n}
\}
\nn\\ &=&
\exp\Bigl\{
-iw\tdlPhi(0)-\pi i \tdlw w
+\frac{i}{2\pi}
\int_0^{2\pi}\Phi(\wssig)\tdlPhi'(\wssig) d\wssig
\Bigr\}\,.
\label{eqn:Kfb}
\eear
(The first term on the second line 
is required to make the entire expression independent of
the choice of origin on the $\wssig$ direction.)
This can be checked by noting that this map acts
on operators as
$$
\px{\wssig}\Phi(\wssig_0)\rightarrow
-2\pi i\frac{\delta}{\delta\Phi(\wssig_0)}
\,,\qquad
-2\pi i\frac{\delta}{\delta\Phi(\wssig_0)}\rightarrow
-\px{\wssig}\Phi(\wssig_0)\,.
$$

Note, however, that we have the freedom
to multiply the operator by an arbitrary function
of the conserved charges, which are the winding number $w$
and the momentum $p\equiv -i\partial/\partial\phi_0.$
The latter has the following interpretation.
Consider first an exponential function $e^{i p a}$,
where $a$ is some constant. This function acts by shifting
$\phi_0\rightarrow\phi_0+a$, and the effect on $\TKer$
in \eqref{eqn:Kfb} is to replace every $\phi_0$ with
$(\phi_0+a).$ Now, if we average this over various $a$'s
with some weight function $f(a)$, the effect on $\TKer$ would be
to multiply it by some function of $\tdlw.$
Therefore, the ambiguity in $\TKer$ can be rephrased as the freedom
to multiply by an arbitrary phase that depends on $(w,\tdlw)$
alone. We will ignore this ambiguity and take
\eqref{eqn:Kfb} as the expression that defines the duality kernel.

Based on what we learned in \secref{subsec:Metric}, 
we expect that the T-duality kernel
$\TKer$ defines a topological theory in one less dimension
when we equate the original variables $w$ and $\phi_n$ to their
dual partners $\tilde{w}$ and $\tilde{\phi}_n$.
Setting $\tdlw=w$ and $\tdlphi_n=\phi_n$
in \eqref{eqn:Kfb}, we find that the diagonal of $\TKer$
is identically zero---this is a trivial topological theory.
Note also that the second line of \eqref{eqn:Kfb}
is a
topological expression (independent of the 0+1D metric).
Interpreting $\wssig$ as time,
the discussion above implies that the Hilbert space of our topological theory has only one state.
We now switch the role of $\wssig$ and $\wstau$,
and from now on, unless otherwise stated, we interpret $\wssig$
as spatial and $\wstau$ as temporal.

\subsection{$T^d$ target space}
\label{subsec:Td}
Our first nontrivial (yet simple) example is a $\sigma$-model
with target space $T^2$ that is a product of two circles
$S^1\times S^1$, one with radius $R_1$ and the other with
radius $R_2=1/R_1.$
The action is
\be\label{eqn:ActST2}
\ActS = \sum_{k=1}^2\frac{1}{4\pi R_k^2}\int\left\{
(\px{\wstau}\Phi_k)^2-(\px{\wssig}\Phi_k)^2
\right\}d\wssig d\wstau
\,,
\ee
and the theory is selfdual under a simultaneous T-duality in both directions,
combined with an exchange of the two $S^1$'s.
In order to get a nontrivial result, it turns
out that we need to add to the twist
a reflection in one of the $S^1$'s.
With this reflection, the combined duality
also preserves the complex structure of the $T^2.$

For the Schr\"odinger formalism, we expand
$$
\Phi_k(\wssig) = w^{(k)}\wssig
+\sum_{n=-\infty}^\infty\phi^{(k)}_n e^{i n\wssig}
\,,\qquad k=1,2,
$$
where $\phi^{(k)}_0$ are real and periodic with period $2\pi.$
A state in the Hilbert space is described by a formal wavefunction
$\Psi(w^{(1)}, w^{(2)},\{\phi^{(k)}_n\})$, and T-duality acts as
\bear
\Psi(\{w^{(k)}\},\{\phi^{(k)}_n\}) &\rightarrow&
\tdlPsi(\tdlw^{(k)},\{\tdlphi^{(k)}_n\})
\nn\\ &
=&\sum_{w^{(1)},w^{(2)}}\int\prod_n d\phi^{(k)}_n
 \TKer(\{\tdlw^{(k)}\},\{\tdlphi_n\};
\{w^{(k)}\},\{\phi_n\})\Psi(\{w^{(k)}\},\{\phi^{(k)}_n\})
\,,
\nn
\eear
where the duality kernel 
$\TKer=\TKer(\{\tdlw^{(k)}\},\{\tdlphi_n\};
\{w^{(k)}\},\{\phi_n\})$ is given by
\begin{eqnarray}
\TKer
=&\exp&\Bigl\{
i (\tdlw^{(2)}\phi^{(1)}_0-w^{(1)}\tdlphi^{(2)}_0)
-i (\tdlw^{(1)}\phi^{(2)}_0+w^{(2)}\tdlphi^{(1)}_0)\nn\\
&&+\sum_{n=-\infty}^\infty n \phi^{(1)}_n\tdlphi^{(2)}_{-n}
-\sum_{n=-\infty}^\infty n \phi^{(2)}_n\tdlphi^{(1)}_{-n}
\Bigr\}
\nn\\ =&
\exp&\Bigl\{
-iw^{(1)}\tdlPhi_2(0)-\pi i \tdlw^{(2)} w^{(1)}
+iw^{(2)}\tdlPhi_1(0)+\pi i \tdlw^{(1)} w^{(2)}\nn\\
&&+\frac{i}{2\pi}
\int_0^{2\pi}\tdlPhi_2\px{\wssig}\Phi_1 d\wssig
-\frac{i}{2\pi}
\int_0^{2\pi}\tdlPhi_1\px{\wssig}\Phi_2 d\wssig
\Bigr\}\,.
\label{eqn:Kfb2}
\end{eqnarray}

Now we set $w^{(k)}=\tdlw^{(k)}$ and $\Phi_k=\tdlPhi_k$
in \eqref{eqn:Kfb2}, and find that the diagonal of the duality kernel becomes a topological action
\be\label{eqn:gqT2A}
\ActS =
\frac{1}{2\pi}\int_0^{2\pi}
  \epsilon^{kl}\Phi_k\px{\wssig}\Phi_l d\wssig
-\frac{1}{2\pi}\epsilon^{kl}\Phi_k(0)
  \int_0^{2\pi}\px{\wssig}\Phi_l d\wssig
\,.
\ee
where $\epsilon^{12}=-\epsilon^{21}=1$
and $\epsilon^{11}=\epsilon^{22}=0.$
We then treat $\wssig$ as a time coordinate, which turns
\eqref{eqn:gqT2A} into a 0+1D action.
The second term on the right-hand side is nonlocal,
but is required in order to make the integral independent of
the choice of origin.
If desired, we can eliminate this term
by replacing the integration range $0\le\wssig<2\pi$
with $-\infty<\wssig<\infty$ and taking as a boundary condition
$\Phi_k(-\infty)=0.$
The resulting action is then simply
\be\label{eqn:gqT2}
\ActS =
\frac{1}{2\pi}\int_{-\infty}^\infty
  \epsilon^{kl}\Phi_k\px{\wssig}\Phi_l d\wssig
\,.
\ee
This action describes geometric quantization of the
target space $T^2$ with a symplectic form
$$
\omega = \frac{2}{(2\pi)^2} d\phi_1\wedge d\phi_2\,,
$$
where $0\le\phi_1<2\pi$ and $0\le\phi_2<2\pi$ are
coordinates on $T^2.$
With this symplectic form, the area of the target space is $2$
and there are therefore two quantum states in the Hilbert space
of this simple topological 0+1D theory.
In \secref{subsec:AltCount} we will present a more geometrical
description of these two states, and the distinction between them.

Let us now generalize the discussion to a torus $T^d$ of an arbitrary 
{\it even} dimension $d$, with an arbitrary flat metric
$G_{IJ}d\phi^I d\phi^J$ and antisymmetric $B$-field
$B_{IJ}d\phi^I\wedge d\phi^J$ ($I,J=1,\dots,d$).
For the duality twist, we pick an element
$\zt$ in the duality group $O(d,d,\Z)$ and write it in block
form as
$$
\zt =
\begin{pmatrix} \xA & \xB \\ \xC & \xD \\ \end{pmatrix}
\in O(d,d,\Z)\,,
$$
where $\xA,\xB,\xC,\xD$ are $d\times d$ matrices.
The action on $G_{IJ}$ and $B_{IJ}$ is conveniently expressed
as follows (see \cite{Giveon:1994fu} for a review).
Define the $d\times d$ matrix $E$ by
$$
E_{IJ} = G_{IJ} + B_{IJ}\,.
$$
Then $\zt$ acts as
$$
E\rightarrow (\xA E +\xB)(\xC E +\xD)^{-1}\,.
$$
For our purposes, we pick a selfdual background for which
$$
E = (\xA E +\xB)(\xC E +\xD)^{-1}\,.
$$

The duality acts on the left-moving and right-moving
free fields of the $\sigma$-model as
\be\label{eqn:dualityTd}
\px{-}\Phi\rightarrow
(\xD - \xC E^t)^{-1}\px{-}\Phi
\,,\qquad
\px{+}\Phi\rightarrow
(\xD + \xC E)^{-1}\px{+}\Phi
\,,
\qquad
\px{\pm}\equiv\px{\wstau}\pm\px{\wssig}
\,,
\ee
where $\Phi$ here is understood as a $d$-component vector.
(See (2.4.36) of \cite{Giveon:1994fu}.)
{}From this action we calculate
\bear
\px{\wstau}\Phi
=\frac{1}{2}(\px{+}\Phi+\px{-}\Phi)
&\rightarrow &
\frac{1}{2}((\xD + \xC E)^{-1}\px{+}\Phi
 +(\xD - \xC E^t)^{-1}\px{-}\Phi)
= \mathbf{U}\px{\wstau}\Phi + \mathbf{V}\px{\wssig}\Phi\,,
\nn\\
\px{\wssig}\Phi
=\frac{1}{2}(\px{+}\Phi-\px{-}\Phi)
&\rightarrow &
\frac{1}{2}((\xD + \xC E)^{-1}\px{+}\Phi
 -(\xD - \xC E^t)^{-1}\px{-}\Phi)
= \mathbf{U}\px{\wssig}\Phi + \mathbf{V}\px{\wstau}\Phi
\,,
\nn
\eear
where
\bear
\mathbf{U}&=&(\xD + \xC E)^{-1}(\xD+\xC B)(\xD - \xC E^t)^{-1}\,,\nn\\
\mathbf{V}&=&
-(\xD + \xC E)^{-1}\xC G(\xD - \xC E^t)^{-1}\,.\nn
\eear

Taking the range $-\infty<\wssig<\infty$
and boundary conditions $\Phi(-\infty)=0$,
to avoid complications, we get (up to total derivatives) the T-duality kernel
\bear
\TKer = \exp\Bigl\{
\frac{i}{4\pi}\int_{-\infty}^\infty
\Bigl(
\Phi (2\mathbf{X})\px{\wssig}\Phi
+\tdlPhi (2\mathbf{Y})\px{\wssig}\tdlPhi
+\tdlPhi (2\mathbf{Z})\px{\wssig}\Phi
\Bigr)d\wssig
\Bigr\}\,,
\label{eqn:TKerTd}
\eear
where
\be
2\mathbf{X}=-B+G\mathbf{V}^{-1}\mathbf{U}\,,\quad
2\mathbf{Y}=B+G\mathbf{U}\mathbf{V}^{-1}\,,\quad
\mathbf{Z}^t=G\mathbf{V}^{-1}\,.\label{tsol}
\ee
This is found by solving\footnote{In manipulating the matrices here and below, 
it is essential to note the following properties of $O(d,d,\Z)$ matrices: 
$\xA^t\xC+\xC^t\xA=0$, $\xB^t\xD+\xD^t\xB=0$, $\xA^t\xD+\xC^t\xB=1$.}
\bear
\left(-2\pi iG^{-1}\frac{\delta}{\delta\tdlPhi}-G^{-1}B\px{\wssig}\tdlPhi\right)\TKer
& = &
\mathbf{U}\left(2\pi iG^{-1}\frac{\delta}{\delta\Phi}
 -G^{-1}B\px{\wssig}\Phi\right)\TKer+\mathbf{V}\px{\wssig}\Phi\TKer
  \,,\nn\\
\px{\wssig}\tdlPhi\TKer
& = & \mathbf{U}\px{\wssig}\Phi\TKer+\mathbf{V}\left(2\pi iG^{-1}
 \frac{\delta}{\delta\Phi}-G^{-1}B\px{\wssig}\Phi\right)\TKer
 \,.\nn
\eear
After a little algebra, \eqref{tsol} can be simplified as
\be
2\mathbf{X}=-E+(\xC^{t})^{-1}(\xD+\xC E)^{-1}\,,\quad 2\mathbf{Y}=-\xC^{-1}\xD\,,\quad
\mathbf{Z}=\xC^{-1}\,.
\ee

Setting $\Phi=\tdlPhi$ in \eqref{eqn:TKerTd}, we get
the topological action
\be\label{eqn:gqTd}
\ActS =
\frac{1}{4\pi}\int_{-\infty}^\infty
\Phi(2\mathbf{W})\px{\wssig}\Phi
d\wssig
\,,
\ee
where
\bear
2\mathbf{W}&=&2\mathbf{X}+2\mathbf{Y}+\mathbf{Z}-\mathbf{Z}^{t}
 = (\xC^t)^{-1}(\xD+\xC E)^{-1}
 - \xC^{-1}(\xD+\xC E)+ \xC^{-1} - (\xC^t)^{-1}\,.\nn
\eear
This describes geometric quantization of the target space
$T^d$ with the symplectic form given by
$$
\omega = \frac{\mathbf{W}_{IJ}}{(2\pi)^d}d\phi^I\wedge d\phi^J\,.
$$
For example, we can recover the previous case with $T^2$ target space by setting
$$
G=\begin{pmatrix} R_1^2 & 0 \\ 0 & R_1^{-2} \end{pmatrix}\,,\quad
B=0\,,\quad \xA = \xD = 0\,,\quad \xB=\xC = J \equiv
\left(\begin{array}{rr}
 0 & -1 \\ 1 & 0 \\ \end{array}\right)\,.
$$
The matrix $J$ is so chosen as to incorporate the exchange of 
two circles and reflection in one of them.
The number of states in the Hilbert space is
$$
n = \mathrm{Pf}(2\mathbf{W})\,,
$$
where $\mathrm{Pf}$ denotes the Pfaffian.

Let us specialize again to the case of $T^2.$
In the study of T-duality for $T^2$, one usually
defines the complex combination
$$
\rho = B_{12} + i\sqrt{\det G}\,.
$$
The duality group $O(2,2,\Z)$ is essentially two copies
of $\SL(2,\Z)$, one acting on $\rho$ and the other acting
on the complex structure of $T^2$ in a geometrical way.
Under an element
$$
\begin{pmatrix} \xa & \xb \\ \xc & \xd \\ \end{pmatrix}
$$
in the first $\SL(2,\Z)$ factor,  $\rho$ transforms as
as
$$
\rho\rightarrow \frac{\xa\rho+\xb}{\xc\rho+\xd}\,.
$$
We now have three possibilities for
a duality twist $\zt$, which are analogous
to the list in \secref{subsec:Stwist}:
\begin{enumerate}
\item
$\rho=i$ and $\zt=\left(\begin{array}{rr}
 0 & -1 \\ 1 & 0 \\ \end{array}\right)$,
with $2$ vacua;
\item
$\rho=e^{\pi i/3}$ and
$\zt=\left(
\begin{array}{rr} 1 & -1 \\ 1 & 0 \\ \end{array}
\right)$,
with $1$ vacuum;
\item
$\rho=e^{\pi i/3}$ and
$\zt=\left(
\begin{array}{rr} -1 & 1 \\ -1 & 0 \\ \end{array}
\right)$,
with $3$ vacua.
\end{enumerate}

\subsection{An alternative way of counting vacua}
\label{subsec:AltCount}
We will now describe a more geometrical interpretation
for the vacua of the topological 0+1D theories that we obtained
in \secref{subsec:Td}.
We will concentrate on the simple $T^2$ target space
with action \eqref{eqn:ActST2}.
The trick is simple: perform T-duality only on one of
the two circles, say the one corresponding to $\Phi_2.$
We now have two circles of equal radius $R_1$
and a target space $T^2$ with complex
structure $\tau=i.$
In this picture selfduality is a geometrical isometry of $T^2$,
which in the realization of $T^2$ as a lattice
$\C/(\Z+\tau\Z)$ corresponds to rotation of $\C$ by $\pi/2.$
A similar duality can be applied for the other selfdual
values of $\rho$ (which becomes the complex structure $\tau$ 
after T-duality only on one circle) from the list at the end of \secref{subsec:Td}.
We will proceed with a general $\tau.$

We take the 1+1D coordinates to be $(\wssig,\wstau)$,
and the twist will be at $\wssig=0\sim 2\pi.$
Set
\be\label{eqn:XPhi}
\XPhi(\wssig,\wstau)\equiv
\Phi^1(\wssig,\wstau) + \tau\Phi^2(\wssig,\wstau)\,,
\ee
to be the complex field of the 1+1D $\sigma$-model.
The boundary conditions are geometrical:
$$
\XPhi(0,\wstau) = e^{i\pht}\XPhi(2\pi,\wstau)\,,
$$
where
$$
e^{i\pht} \equiv\xc\rho+\xd\,.
$$
The twist has a number of fixed points $z_r$ ($r=1,2,\dots$)
that satisfy
$$
e^{i\pht}z_r - z_r\in\Z+\tau\Z\,,
$$
i.e., rotation by $\pht$ keeps the point on $T^2$
that is parameterized by $z_r$ invariant.
The number of fixed points is as follows:
\begin{enumerate}
\item
For $\rho=i$ and $\pht=\pi/2$,
we have $r=1,2$ and the fixed points are
$z_1=0$ and $z_2=(1+i)/2$;
\item
For $\rho=e^{\pi i/3}$ and $\pht=\pi/3$
we have only one fixed point
$z_1 = 0$;
\item
For $\rho=e^{\pi i/3}$ and $\pht=2\pi/3$
we have $3$ fixed points
$z_1 = 0$, $z_2=(\rho+1)/3$, and $z_3=2(\rho+1)/3.$
\end{enumerate}
Each fixed point $z_r$ defines a different topological
sector of the $\sigma$-model via the mode expansion
$$
\XPhi = z_r
+\sum_{q\in\Z+\frac{\pht}{2\pi}}
\frac{1}{q}\alpha_q e^{-i q(\wstau+\wssig)}
+\sum_{q\in\Z-\frac{\pht}{2\pi}}
\frac{1}{q}\widetilde{\alpha}_q e^{-i q(\wstau-\wssig)}
\,.
$$
The modes are all fractional, and there are no zero modes here.
The number of ground states is therefore the number of topological
sectors, labeled by the index $r.$ The result, which is listed above,
agrees with the result listed at the end of \secref{subsec:Td}.

\subsection{K\"ahler $\sigma$-models and the Witten index}
\label{subsec:Kahler}
The discussion in the previous subsections can be extended
to supersymmetric $\sigma$-models.
We will encounter in \secref{sec:Nonabelian} the following situation:
a selfdual
nonlinear 1+1D $\sigma$-model with \SUSY{(2,2)} supersymmetry
compactified on $S^1$ with a duality twist augmented
by an isometry of the K\"ahler target space.
The duality is mirror symmetry \cite{Witten:1991zz}
-\cite{Morrison:1995yh}, and the complex K\"ahler
moduli are at a selfdual value.
We denote the target space by $\TgSp.$

The question is whether the low-energy description (energy
scale much lower than the Kaluza-Klein scale of the
circle compactification) is a topological theory.
We assume that the combination of mirror symmetry and isometry
twist preserves some amount of supersymmetry (half of the
SUSY generators, generally), and
we wish to count the number of vacua,
or at least calculate the Witten index.
\\[0pt]
\indent
In our application below, $\TgSp$ will be the Hitchin's moduli space
$\MH$ (to be reviewed in \secref{subsec:HSpace}),
which is actually hyper-K\"ahler, and not just K\"ahler
(and the $\sigma$-model therefore starts out with \SUSY{(4,4)}
supersymmetry in 1+1D).
But for the time being it is good to start with
a simple prototypical example: $\TgSp=T^2\times\C$
(which also happens to be hyper-K\"ahler)
where the complexified K\"ahler class $\rho$ of $T^2$
is one of the two choices from the end of \secref{subsec:Td}.
The twist along $S^1$
is a combination of mirror symmetry and isometry.
Mirror symmetry is just the T-duality of $T^2$ in this context,
and is described by an element $\zt\in\SL(2,\Z)$
which we also pick out from the list
at the end of \secref{subsec:Td}.
We combine it with a rotation of $\C$ by some nonzero angle $\beta.$
It is not hard to check that
none of the fermionic fields of the $\sigma$-model have zero-modes,
and so it is clear from the discussion above
that the low-energy theory is equivalent to
geometric quantization of the isometry-invariant subspace
of $\TgSp$, which is $T^2\times\{0\}$ (where $\{0\}$
stands for the origin of $\C$).
\\[0pt]
\indent
To see this in more detail,
let us rederive the results of the previous subsections
in this supersymmetric context, using a technique that
will be useful in the more complicated case of $\MH$ later on.
Since none of the fermionic fields have zero modes,
we can count the number of vacua by calculating the Witten index
of the theory. For this purpose, we compactify
time on a (Euclidean) circle $0\le \wstau\le 2\pi T$
and calculate the partition function.
The fermions have periodic boundary conditions that
preserve supersymmetry.
We also introduce
a topological twist \cite{Witten:1991zz}
that turns the supersymmetric \SUSY{(2,2)} $\sigma$-model
into either the A-model or the B-model, which we will
discuss separately below.
Since we are working on a flat worldsheet, the topological
twist has no effect on the partition function.
\\[0pt]
\indent
Since the T-duality twist $\TKer$ commutes with the BRST operator
of  either A-model or B-model,
we can reduce the calculation of the partition
function to a trace of the reduction of $(-1)^F\TKer$
in the finite dimensional Hilbert space of the A-model compactified
on $S^1$ (the $\wstau$-direction),
where $F$ is the fermion number.
The fact that this gives the same result as in \secref{subsec:AltCount}
can be understood as a variant of the
Lefschetz--Hopf fixed-point theorem which relates the number
of fixed points (counted with multiplicity) of a continuous
map on a manifold to the trace of the induced map on cohomology,
and can be derived from a topological field theory
\cite{Schwarz:1989fb}\cite{Niemi:1993vp}\cite{Li:2005eh}.
Let us now proceed to the details.

\subsubsection*{A-model}
For a $T^2$ target space as in \secref{subsec:AltCount},
the A-model action is:
\bear
L &=& \frac{4\pi\Imx\rho}{\Imx\tau} \int\left(
\frac{1}{2}\bpx{\bz}\bXPhi\px{z}\XPhi
+\frac{1}{2}\bpx{\bz}\XPhi\px{z}\bXPhi
+i\psi_z\bpx{\bz}\chi
+i\bpsi_\bz\px{z}\bchi\right) d^2z
\nn\\ &&
+\frac{2\pi\Rex\rho}{\Imx\tau}\int\left(
\bpx{\bz}\bXPhi\px{z}\XPhi
-\bpx{\bz}\XPhi\px{z}\bXPhi
\right)d^2z
\label{eqn:AcVT2}
\,,
\eear
where $z=\wssig+i\wstau$,
$\XPhi$ is the same complex coordinate on the target
space as in \eqref{eqn:XPhi},
$\psi_z,\bpsi_\bz,\chi,\bchi$ are fermionic fields,
$\rho$ is the (complex) K\"ahler modulus of $T^2$
(taken from the list
at the end of \secref{subsec:Td}), and
$\tau$ is the complex structure of $T^2$, which decouples
from the topological theory.
The BRST symmetry acts as \cite{Witten:1991zz}:
\be\label{eqn:BRST}
\delta\XPhi = i\epsilon\chi\,,\quad
\delta\bXPhi = i\epsilon\bchi\,,\quad
\delta\chi=\delta\bchi = 0\,,\quad
\delta\psi_z=-\epsilon\px{z}\bXPhi\,,\quad
\delta\bpsi_\bz=-\epsilon\bpx{\bz}\phi\,.
\ee

The Hilbert space of the topological A-model compactified
on $S^1$ (to be understood as the $\wstau$ direction, according
to the discussion above) is in one-to-one correspondence
with the Dolbeault cohomology of $T^2.$
A basis of
local BRST-cohomology operators which correspond to these
states consists of \cite{Witten:1991zz} 
$
1,\,
\chi,\,
\bchi,\,
\bchi\chi,
$
which correspond to the following
representatives of the Dolbeault cohomology
of $T^2$:
$
1,\,
d\ZXPhi,\,
d\bZXPhi,\,
d\bZXPhi\wedge d\bZXPhi.
$
(Here $\ZXPhi,\bZXPhi$ are coordinates on $T^2$
which are in one-to-one correspondence with
the $\sigma$-model fields $\XPhi,\bXPhi.$)

The T-duality element $\TKer$ acts on the fermionic fields of
the A-model as follows (compare with \eqref{eqn:dualityTd}):
\be\label{eqn:Ttopo}
\chi\rightarrow e^{i\pht}\chi
\,,\quad
\bchi\rightarrow e^{i\pht}\bchi
\,,\quad
\psi\rightarrow e^{-i\pht}\psi
\,,\quad
\bpsi\rightarrow e^{-i\pht}\bpsi
\,,
\ee
and commutes with the BRST transformation \eqref{eqn:BRST}.
The T-duality element $\TKer$
therefore acts on an A-model operator that corresponds
to a $(p,q)$-Dolbeault cohomology class as multiplication
by the phase $e^{i(p+q)\pht}.$
The action depends only on the total degree of the form,
as it should, since the A-model is independent of the complex
structure of the target space.

Using the state-operator correspondence, we can now determine
the action of $\TKer$ on states, up to a phase.
Letting $\ket{1}$ be the state corresponding to the operator $1$,
the phase is $\bra{1}\TKer\ket{1}.$
The Witten index is then
$$
I = \tr\{(-1)^F\TKer\} = (1-e^{i\pht})^2\bra{1}\TKer\ket{1}.
$$
Thus, we get
$$
|I| = |1-e^{-i\pht}|^2 = 2(1-\cos\pht).
$$
(And the missing phase is $\bra{1}\TKer\ket{1}=\pm e^{-i\pht}.$)
This agrees with the results of \secref{subsec:AltCount}.

\subsubsection*{B-model}
The B-model action with $T^2$ target space is
\bear
L &=& \frac{4\pi\Imx\rho}{\Imx\tau}\int_\Sigma d^2z\Bigl(
\frac{1}{2}\px{z}\XPhi\px{\bz}\bXPhi
+\frac{1}{2}\px{z}\bXPhi\px{\bz}\XPhi
+\frac{i}{2}\eta(\px{z}\rho'_\bz+\px{\bz}\rho'_z)
+\frac{i}{2}\theta(\px{\bz}\rho'_z-\px{z}\rho'_\bz)
\Bigr)\,.
\nn
\eear
with the BRST action
\bear
\delta\XPhi = 0,\quad
\delta\bXPhi = i\epsilon\eta,\quad
\delta\eta =\delta\theta = 0,\quad
\delta\rho' = -\epsilon d\phi\,.
\eear
The BRST-invariant operators are
$
1,
\eta,
\theta,
\eta\theta,
$
which correspond to the following elements of $H^p(\wedge^q T^{(1,0)}(T^2))$:
$1,d\bXPhi,\ppx{\XPhi},d\bXPhi\ppx{\XPhi}.$

T-duality acts as
$$
\eta\rightarrow \eta\cos\pht+i\theta\sin\pht\,,\quad
\theta\rightarrow i\eta\sin\pht+\theta\cos\pht\,,\quad
\rho'_z\rightarrow e^{i\pht}\rho'_z\,,\quad
\rho'_\bz\rightarrow e^{-i\pht}\rho_\bz\,,
$$
and we can verify that $\tr\{(-1)^F\TKer\} = 2-2\cos\pht.$

\section{Analysis for $U(1)$ Super-Yang--Mills}
\label{sec:Abelian}
We now study 3+1D Yang--Mills theory with an S-duality twist
for the case of a $U(1)$ gauge group.
In this case, there is an exact expression for the S-duality
kernel, which is well-known, and it is straightforward
to find the topological 3D theory associated with the
S-duality twist.

\subsection{The duality kernel for $U(1)$ Yang--Mills theory}
\label{subsec:KerU1}
We take pure $U(1)$ Yang--Mills theory with 1-form gauge field
$A$ defined on $\Mf_3.$
The S-duality kernel
$\SKer(A,\tA)$ acts on the wavefunction $\Psi\{A\}$ representing
a state so that
$$
\widetilde{\Psi}\{A\}\equiv\int [\cD\tA]\SKer(A,\tA)\Psi(\tA)
$$
is the wavefunction of the S-dual state.

For an S-duality transformation we have
$$
\tau\rightarrow\frac{\xa\tau+\xb}{\xc\tau+\xd}\,,
\qquad
E_i\rightarrow \xa E_i + \xb B_i\,,
\quad
B_i\rightarrow \xc E_i + \xd B_i\,.
$$
The action of S-duality in the quantum theory on an
arbitrary manifold was described in \cite{Witten:1995gf}.
A closed expression for the S-duality kernel
appears in 
\cite{Lozano:1995aq}\cite{Witten:2003ya}\cite{Gaiotto:2008ak}:
\be\label{eqn:SAtA}
\ASKer(A,\tA) = \exp\left\{
\frac{i}{4\pi\xc}\int (
\xd A\wedge dA
-2\tA\wedge dA
+\xa\tA\wedge d\tA
)\right\}.
\ee
This is determined by requiring the operator equations
$$
\tE_i\ASKer = \ASKer(\xa E_i + \xb B_i),
\qquad
\tB_i\ASKer = \ASKer(\xc E_i + \xd B_i).
$$
Here we can take $E_i\equiv -2\pi i\delta/\delta A_i$.

Now set $A=\tA$ (up to a gauge transformation)
in \eqref{eqn:SAtA}. We get
$$
\AcV(A)\equiv \frac{\xa+\xd-2}{4\pi\xc}\int A\wedge dA\,.
$$
This is a Chern--Simons theory at level
$\lvk\equiv(\xa+\xd-2)/\xc.$
For a generic $\SL(2,\Z)$ element
$\zs=\begin{pmatrix} \xa & \xb \\ \xc & \xd \\ \end{pmatrix}$
this is not an integer,
but for the special values $\zs=\zs',-\zs',\zs'',-\zs''$
we get integral levels $\lvk=-2,2,-1,3$, respectively.

\subsection{Low-energy limit of an $\SL(2,\Z)$-twisted compactification}
\label{subsec:LEtwisted}
Now, let us compare the Chern--Simons
action that we obtain from the diagonal $\ASKer(A,A)$
to the action that we obtain from compactifying
$U(1)$ Yang--Mills theory on $S^1$ of radius $R$
with an $\zs$-twist, in the limit $R\rightarrow 0$,
as in \secref{subsec:Stwist}.

Let us describe the full action in detail.
We assume that the $\zs$-twist is at $x_3=0\simeq 2\pi R.$
The Yang--Mills field $A(x_0,x_1,x_2,x_3)$ is defined
in the range $0\le x_3\le 2\pi R$ {\it without}
imposing periodic boundary conditions.
The Yang--Mills coupling constant is either $\tau=i$
or $\tau=e^{\pi i/3}$, according to whether $\zs=\zs'$
or $\zs=\pm\zs''.$ As is customary, we set $\tau\equiv\tau_1+i\tau_2.$
We also denote
$$
A'\equiv A(x_0,x_1,x_2,x_3=0)
\,,\qquad
A''\equiv A(x_0,x_1,x_2,x_3=2\pi R)
\,.
$$
The full action is
$$
\AcV = \AcV_{YM} + \AcV_{X}\,,
$$
where $\AcV_{YM}$ is the bulk Yang--Mills action
$$
\AcV_{YM} \equiv
\int_{x_0, x_1, x_2}\int_{x_3=0}^{2\pi R}
\left(\frac{1}{2\gYM^2}\, F\wedge {}^*F + \frac{\theta}{4\pi^2}\,F\wedge F\right)\,,
$$
and $\AcV_{X}$ consists of ``boundary terms''
$$
\AcV_{X}\equiv
\frac{1}{4\pi}\int_{x_0, x_1, x_2} \LagJ\,,
$$
where
$$
\LagJ\equiv
\frac{1}{\xc}\left(
\xd A'\wedge dA''
-2 A'\wedge dA''
+\xa A''\wedge dA''
\right)
$$
is the integral of a gauge-invariant expression:
$$
d\LagJ =
\frac{1}{\xc}(
\xd F'\wedge F''
-2 F'\wedge F''
+\xa F''\wedge F''
)\,.
$$

The equations of motion are Maxwell's equations in the bulk,
but with boundary conditions:
$$
F'' = (\xc\tau_1+\xd)\,F' + \xc\tau_2\,{}^*F'\,,
\qquad
F' = -(\xc\tau_1-\xa)\,F'' -\xc\tau_2\,{}^*F''\,.
$$
These two conditions are equivalent for selfdual values of
$\tau$ and corresponding $\zs$.

Define the complex-valued 2-forms
$$
F_{\pm}'\equiv {}^*F' \pm i F'\,,\qquad
F_{\pm}''\equiv {}^*F'' \pm i F''\,.
$$
Then, the boundary conditions can be written as
\be\label{eqn:bcpm}
F_{+}'' = (\xc\tau +\xd) F_{+}'\,,
\qquad
F_{-}'' = (\xc\overline{\tau} +\xd) F_{-}'\,.
\ee
As noted earlier, $|\xc\tau+\xd|=1$ so that we can write
$\xc\tau+\xd = e^{2\pi i\fx}$ with
$$
\fx = \left\{\begin{array}{ll}
\frac{1}{4} & \text{for $\tau=i$ and $\zs=\zs'$,} \\
\frac{1}{6} & \text{for $\tau=e^{\pi i/3}$ and $\zs=\zs''$,} \\
\frac{2}{3} & \text{for $\tau=e^{\pi i/3}$ and $\zs=-\zs''$.}
\end{array}\right.
$$
From the boundary conditions \eqref{eqn:bcpm}
we find the Fourier mode decomposition
$$
F_\pm =
\sum_{j\in\Z} e^{\frac{i (j +\fx) x_3}{R}}
f_{j+\fx}^{(+)}(x_0, x_1, x_2)\,.
$$
Because $j+\fx$ is never zero, we see that the fields
$f_{j+\fx}^{(+)}$ are massive in 2+1D with masses given by
$|j+\fx|/R.$
The classical analysis, however,
cannot tell us the multiplicity of the vacuum.
But since the low-energy description is a Chern--Simons
theory at level $\lvk\equiv(\xa+\xd-2)/\xc$
we expect to get a multiplicity of $\lvk^h$ vacua
when formulated on a compact genus-$h$
Riemann surface $\RSC_h.$

\subsection{Supersymmetry}
\label{subsec:IncludeSUSY}
Now let us extend the discussion to
a free vector multiplet of \SUSY{4} SYM.
The extra fields are free scalars and fermions.
We need to impose the boundary conditions \eqref{eqn:phipsibc}
combined with the $\zs$-twist.
The action of $\zs$ on the $6$ scalar fields of the vector
multiplet is trivial, since it commutes with the $SO(6)$ R-symmetry.
One might consider the possibility of $\zs$ acting as an overall
$(-)$ sign, which corresponds to the nontrivial element
in the center of $SO(6)$, but this is a matter of definition,
and we can always absorb it in the R-symmetry twist $\gtw.$
We then get
2+1D scalar Klauza-Klein modes with  masses
$(j+(\ftw_a+\ftw_b)/2\pi)/R$ ($1\le a<b\le 4$),
where $j\in\Z$ and $\ftw_a$ are as in \eqref{eqn:gtwgen}.

Now consider the free fermions of the \SUSY{4} vector multiplet.
By \eqref{eqn:zsQ} and \eqref{eqn:phipsibc},
their boundary conditions are
$$
\psi^{\a a}(x_3=2\pi R) =
e^{\frac{i}{2}\pht+i\ftw_a}\psi^{\a a}(x_3=0)
\,,\qquad
\bpsi_a^{\dta}(x_3=2\pi R) =
e^{-\frac{i}{2}\pht-i\ftw_a}\bpsi_a^{\dta}(x_3=0)\,.
$$
This gives
2+1D fermionic Klauza-Klein modes with  masses
$(j+(\ftw_a+\tfrac{1}{2}\pht)/2\pi)/R$ ($1\le a\le 4$).

For a generic choice of $\gtw$ (i.e., generic $\ftw_a$)
there are neither fermionic nor bosonic zero modes.
This is also the case for the \SUSY{6} supersymmetric $\gtw$
in \eqref{eqn:gtwN=6}.
For the \SUSY{4} supersymmetric choices of $\gtw$ in
\eqref{eqn:gtwN=4} there are no zero modes unless
the phase $\ftw_4$ is chosen so that
$e^{i(\tfrac{1}{2}\pht+\ftw_4)}=1.$
In that case the subgroup of $SU(4)_R$
that commutes with $\gtw$ is
$(SU(2)\times SU(2)\times U(1))/\Z_2.$
The surviving supercharges transform
in the representation
$(\rep{2},\rep{1})_{+1}\oplus(\rep{2},\rep{1})_{-1}$
so the $U(1)$ factor and the leftmost $SU(2)$ factor can
be considered an R-symmetry of the resulting theory,
while the right $SU(2)$ factor is a flavor symmetry.
The low-energy theory comprises of $4$ massless scalar fields
in the representation $(\rep{2},\rep{2})_0$ of
the unbroken R-symmetry, and $4$ massless fermions
in the representation
$(\rep{1},\rep{2})_{+1}\oplus (\rep{1},\rep{2})_{-1}.$
These combine to a 2+1D hypermultiplet.
The moduli space is $\R^4.$

So far we discussed the physical low-energy theory.
Now, let us discuss the action defined by \eqref{eqn:AcV}.
The duality kernel is given by
$$
\SKer(\lV,\sdlV) = \ASKer(A,\sdA)
\delta(\sdPhi-\Phi^\gtw)
\delta(\sdpsi-e^{\frac{i}{2}\pht}\psi^\gtw)
\,,
$$
where $\ASKer(A,\tA)$ is given by \eqref{eqn:SAtA}.
Setting $\sdlV=\lV$ we get, up to an infinite normalization factor,
\be\label{eqn:SKerlVlV}
\SKer(\lV,\lV) = e^{\frac{i k}{4\pi}\int A\wedge dA}
\delta(\Phi)\delta(\psi).
\ee
The normalization factor is, formally, a product
of the determinants (one determinant for each spacetime point
$x$), since
$$
\delta(\Phi(x)-\Phi(x)^\gtw)
\delta(\psi(x)-e^{\frac{i}{2}\pht}\psi(x)^\gtw)
=\delta(\Phi(x))\delta(\psi(x))
\frac{\prod_a(1-e^{i(\ftw_a+\tfrac{1}{2}\pht)})}{
\prod_{a<b}(1-e^{i(\ftw_a+\ftw_b)})}\,.
$$
If the constant factor on the right is well-defined,
nonzero, and finite
(i.e., in the absence of fermionic and bosonic zero modes)
the resulting action $\SKer(\lV,\lV)$ is indeed topological,
after regularization.

Note that at low-energy, in the topological
theory, supersymmetry now acts in a trivial way:
all the supersymmetry generators are identically zero.
This is because by \eqref{eqn:SKerlVlV}
we have $\Phi=0$ and $\psi=0$, and the
equations of motion of the Chern--Simons theory also
set $F=0.$ The vanishing of the SUSY generators
immediately implies that the Hamiltonian is identically zero
(since the Hamiltonian is part of the supersymmetry algebra),
which is consistent with the topological nature of the low-energy
theory.

\section{The nonabelian case}
\label{sec:Nonabelian}
We now turn to the nonabelian case.
Our setting is \SUSY{4} $SU(n)$
SYM compactified on $S^1$ of radius $R$
with an R-symmetry twist $\gtw$ and an
$\SL(2,\Z)$-duality twist
$\zs$ at a point on the circle.\footnote{The $SU(n)$ theory
is not selfdual under the full $\SL(2,\Z)$ group, but rather
only under a subgroup known as $\Gamma_0(n)$. This is because
the dual group of $SU(n)$ is its adjoint form $SU(n)/\Z_n$. 
The difference has to do with allowed electric and magnetic fluxes,
which we will address in \secref{subsec:EffectOfFluxes}. For the 
time being, we will ignore this subtlety.}  
We have argued that the low-energy
limit, $R\rightarrow 0$, is described
by a 2+1D topological field theory.
We ask: {\it what is that field theory?}

In \secref{subsec:Conjecture} we present our
conjecture: the low-energy limit can be described
by a Chern--Simons theory at a level
that is determined by the twist.
We then test this conjecture in \secref{subsec:WIndex}
by calculating the Witten index
of the theory compactified (in an appropriate
way that preserves some supersymmetry) on a Riemann surface,
and we compare the result to the number of vacua of Chern--Simons
theory on that Riemann surface.
We now proceed to the details.

\subsection{A conjecture}
\label{subsec:Conjecture}
Our conjecture is as follows.
For the values of $n,\tau,\zs,\pht$ listed below,
the low-energy limit of \SUSY{4} SYM with gauge group $SU(n)$
and complex coupling constant $\tau$, compactified
on $S^1$ with an $\SL(2,\Z)$-twist $\zs$ and R-symmetry
twist $\gtw$ (determined by $\pht$) as in \eqref{eqn:gtwN=6},
is described by a (three-dimensional) pure Chern--Simons
theory with the same gauge group $SU(n)$ and at level $\lvk$
that is given by:
\begin{itemize}
\item
for $\tau=i$, $\pht = \frac{\pi}{2},$
$\zs=\zs'\equiv
\left(\begin{array}{rr}
 0 & -1 \\ 1 & 0 \\ \end{array}\right),$ and $n=1,2,3$,
we have $\lvk=-2$;
\item
for $\tau=e^{\pi i/3}$, $\pht= \frac{\pi}{3},$
$\zs=\zs''\equiv\left(
\begin{array}{rr} 1 & -1 \\ 1 & 0 \\ \end{array}
\right),$ and $n=1,2,3,4,5$, we have $\lvk=-1$;
\item
for $\tau=e^{\pi i/3}$, $\pht= \frac{4\pi}{3},$
$\zs=-\zs''=\left(
\begin{array}{rr} -1 & 1 \\ -1 & 0 \\ \end{array}
\right),$ and $n=1,2$, we have $\lvk=3$.
\end{itemize}
Supersymmetry is realized trivially (all generators are zero).
The levels $\lvk$ are conjectured by extension
from the $U(1)$ case discussed in \secref{subsec:KerU1}.
The restrictions on the rank $n$ are in order to eliminate
zero-modes of scalar fields, as discussed in \secref{subsec:ZModes}.
The negative values of $\lvk$ for the first two cases in the list can, of course,
be flipped to positive values with the help of a parity transformation.

The conjecture implies that the expectation value of a large smooth
Wilson loop can be calculated from Chern--Simons theory.
In Euclidean signature,
let $0\le x_3 < 2\pi R$ be a periodic coordinate on $S^1$,
and let $\LoopC\subset\R^3$ be a loop at a constant $x_3$
(and here $\R^3$ represents
the remaining three dimensions of the problem).
We assume that the curvature of $\LoopC$ is small compared
to $R^{-1}$ and that the loop is not self-intersecting
or ``close'' to being self-intersecting. (More precisely,
we assume that the intersection of $\LoopC$
with any ball in $\R^3$ of radius of the order of $R$
or less is topologically connected.)
The expectation value $\langle W(\LoopC)\rangle$
of a Wilson loop $W(\LoopC)$
is then given, by conjecture, by a similar expectation value
$\langle W(\LoopC)\rangle$ in the corresponding
three-dimensional Chern--Simons theory.
It can therefore be calculated
using the techniques developed in \cite{Witten:1988hf}.

The restriction on the curvature of the loop
can presumably be dropped if we supersymmetrize
the loop, as in \cite{Rey:1998ik}\cite{Maldacena:1998im}.
Since the scalars and fermions are set to zero at low-energy,
by our conjecture, the supersymmetrization should have no effect
on the Chern--Simons side. 

\subsection{Relations among the Chern--Simons levels}
\label{subsec:ConjRel}
The three cases corresponding to the
twists $\zs=\zs',\zs'',-\zs''$ are related,
and were it not for
the different $\gtw$-twists,
a proof of the conjecture for any one of them
would have implied the rest.
To see this, set
$$
T\equiv
\begin{pmatrix} 1 & 1 \\ 0 & 1 \\ \end{pmatrix}
\,,\qquad
S\equiv
\begin{pmatrix} 0 & -1 \\ 1 & 0 \\ \end{pmatrix}
\,.
$$
Then
$$
\zs'=S\,,\qquad
\zs''= T S\,,\qquad
-\zs''=T S^{-1}\,.
$$
The action of $T$ is simple to describe.
It multiplies the wavefunction of 3+1D SYM by
the level $\lvk=1$ Chern--Simons phase, as in
\eqref{eqn:CSshift}.
Furthermore, if the kernel for $S$ is $\SKer(\lV,\sdlV)$,
in the notation of \secref{subsec:Sker},
then the kernel for $S^{-1}$ is
$\SKer(\sdlV,\lV)^{*}$, since $\hat{\SKer}$ is a unitary operator.
It follows that if the diagonal
of the kernel for $\zs'$ corresponds to Chern--Simons theory
at level $\lvk$, then $\zs''$ is described by level
$(\lvk+1)$ and $-\zs''$ by level $(1-\lvk)$ (which happens to
be true from the list of \secref{subsec:Conjecture}).
However, since the R-symmetry twists are different
in the three cases, we do not know how to prove this
relation definitively.

\subsection{Compactification on a Riemann surface $\RSC_h$}
\label{subsec:CompRSC}
\FIGURE{
\begin{picture}(400,150)


\put(195,70){\begin{picture}(200,100)
\put(-3,15){$\RSC_h$}
\put(-43,-45){Riemann surface}
\put(-33,-55){area $\Area$}
\thinlines
\qbezier(0,10)(10,10)(20,20)
\qbezier(20,20)(30,30)(50,30)
\qbezier(50,30)(70,30)(80,20)
\qbezier(80,20)(90,10)(90,0)

\qbezier(0,10)(-10,10)(-20,20)
\qbezier(-20,20)(-30,30)(-50,30)
\qbezier(-50,30)(-70,30)(-80,20)
\qbezier(-80,20)(-90,10)(-90,0)

\qbezier(0,-10)(10,-10)(20,-20)
\qbezier(20,-20)(30,-30)(50,-30)
\qbezier(50,-30)(70,-30)(80,-20)
\qbezier(80,-20)(90,-10)(90,0)

\qbezier(0,-10)(-10,-10)(-20,-20)
\qbezier(-20,-20)(-30,-30)(-50,-30)
\qbezier(-50,-30)(-70,-30)(-80,-20)
\qbezier(-80,-20)(-90,-10)(-90,0)

\qbezier(-75,2)(-65,-8)(-50,-8)
\qbezier(-50,-8)(-35,-8)(-25,2)

\qbezier(-70,0)(-60,10)(-50,10)
\qbezier(-50,10)(-40,10)(-30,0)

\qbezier(75,2)(65,-8)(50,-8)
\qbezier(50,-8)(35,-8)(25,2)

\qbezier(70,0)(60,10)(50,10)
\qbezier(50,10)(40,10)(30,0)
\end{picture}}

\put(320,70){\begin{picture}(20,20)
\thicklines
\put(-5,-5){\line(1,1){10}}
\put(-5,5){\line(1,-1){10}}
\end{picture}}

\put(40,70){\begin{picture}(20,20)
\thinlines
\put(-17,62){S-twist}
\put(-4,40){$S$}
\put(-5,29){\line(1,1){10}}
\put(5,29){\line(-1,1){10}}
\put(-4,24){$\gtw$}
\put(-19,52){\& R-twist}
\qbezier(34,0)(34,14)(24,24)
\qbezier(24,24)(14,34)(0,34)

\qbezier(34,0)(34,-14)(24,-24)
\qbezier(24,-24)(14,-34)(0,-34)

\qbezier(-34,0)(-34,14)(-24,24)
\qbezier(-24,24)(-14,34)(0,34)

\qbezier(-34,0)(-34,-14)(-24,-24)
\qbezier(-24,-24)(-14,-34)(0,-34)

\put(-3,-45){$S^1$}
\put(-23,-55){$0\le x_3 < 2\pi R$}
\end{picture}}

\put(90,70){\begin{picture}(20,20)
\thicklines
\put(-5,-5){\line(1,1){10}}
\put(-5,5){\line(1,-1){10}}
\end{picture}}

\put(390,70){\begin{picture}(20,20)
\thinlines
\qbezier(34,0)(34,14)(24,24)
\qbezier(24,24)(14,34)(0,34)

\qbezier(34,0)(34,-14)(24,-24)
\qbezier(24,-24)(14,-34)(0,-34)

\qbezier(-34,0)(-34,14)(-24,24)
\qbezier(-24,24)(-14,34)(0,34)

\qbezier(-34,0)(-34,-14)(-24,-24)
\qbezier(-24,-24)(-14,-34)(0,-34)

\put(-3,37){$S^1$}
\put (-16,10){(or $\R$)}
\put(-23,-55){$0\le x_4 < 2\pi T$}
\end{picture}}
\end{picture}
\caption{
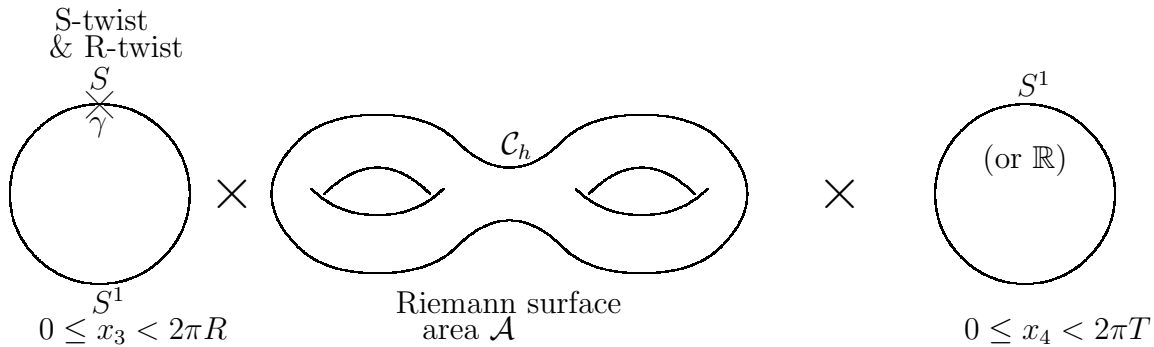
Our setting is \SUSY{4} $SU(n)$ SYM compactified
on an $S^1$ with
an R-symmetry and S-duality twist
times a Riemann surface $\RSC_h$ of genus $h$
($h=2$ in the picture).
The remaining dimension is also compactified on another $S^1.$
The R-symmetry bundle
is nontrivial over $\RSC_h$
so as to preserve half of the supersymmetry.
}
\label{fig:RSxS1xS1}
}
In order to explore the conjecture presented in
\secref{subsec:Conjecture} we wish to find
a topological quantity that can be computed in
\SUSY{4} SYM, using what is already known about the
action of S-duality, and then compare the result to
what our conjecture predicts in terms of Chern--Simons theory.
As a first step, we compactify
the theory on a Riemann surface $\RSC_h$ of genus $h.$
In other words, we consider the theory on 
$X=S^1_R\times\RSC_h\times \R$, where the subscript
$R$ refers to the radius of the circle,
with the $\gtw$ and $\zs$ twists
setting the boundary conditions along $S^1_R$.

We also wish to preserve some amount of supersymmetry,
so that Witten-index techniques could be applicable.
We can do this by turning on an appropriate
(topologically nontrivial) background
gauge field along $\RSC_h$
for the unbroken R-symmetry---an operation
known as ``twisting''
\cite{Witten:1991zz,Bershadsky:1995qy},
which we will briefly review.

For this additional twisting we are only allowed
to use the unbroken subgroup of the R-symmetry group.
The already present
R-twist of \eqref{eqn:gtwN=6} breaks the R-symmetry group
of \SUSY{4} SYM down to $U(3)\subset SU(4)_R$, under which
the $6$ supercharges transform as the sum of the fundamental
and anti-fundamental representations $\rep{3}+\rep{\overline{3}}.$
The supercharges
also transform as a spinor (left-moving plus right-moving)
in the two directions of $\RSC_h$,
which means that one component transforms as a section
of the $SO(2)$ bundle associated with the phase of
the square-root of the canonical line bundle $\CanK$ of $\RSC_h$
and the other component transforms as a section
of the opposite bundle
(the one associated with the anti-canonical bundle $\bCanK$).
For genus $h\neq 1$ these are nontrivial bundles, and
there are therefore no covariantly constant spinors on $\RSC_h$,
and supersymmetry is completely broken.

The procedure of twisting restores supersymmetry
by adding a background $SU(4)_R$ gauge field
that is proportional to the spin connection of $\RSC_h.$
This modifies the covariant derivative of the fermions
and scalars that are charged under $SU(4)_R.$
The spin connection of $\RSC_h$ can be thought of
as a gauge field for the group of rotations $SO(2)$ of
the fibers of the tangent-bundle of $\RSC_h.$
To specify the topological twist we need to specify
an element $\twsu$ in the (R-symmetry) Lie algebra $\frak{su}(4).$
Denoting by $\omega_j$ ($j=z,\bz$) the components
of the spin connection on $\RSC_h$,
the covariant derivative of a left-moving fermion
in the $\RSC_h$ direction $j$
is then $D_j = \px{j}-\frac{1}{2}\omega_j-\omega_j\twsu.$
Here $\twsu$ acts on the R-symmetry indices of the field,
and we assume that it commutes with the R-symmetry twist
in \eqref{eqn:gtwN=6}: $\twsu\in \frak{u}(3)\subset \frak{su}(4).$
In the basis that corresponds to \eqref{eqn:gtwgen} 
we therefore take
\be\label{eqn:twsu}
\twsu\equiv
\begin{pmatrix}
\twx_1 & &  & \\
& \twx_2 &  & \\
& & \twx_3  & \\
& & & -\sum_1^3\twx_i  \\
\end{pmatrix} \in \frak{u}(3)\subset \frak{su}(4)_R\,.
\ee
After the topological twist (and contraction
with the zweibein if necessary), the scalars and fermions
turn into sections of generally nontrivial line bundles
over $\RSC_h$ which are certain powers of $\CanK$
or $\bCanK.$ The supercharges are also sections of such
line bundles, and the number of conserved supersymmetries
is the number of supercharges that transform
in the trivial bundle \cite{Witten:1991zz}.

The supercharges that transform in the $\rep{4}$ of $SU(4)_R$
are also left-moving spinors under the 3+1D Lorentz group,
and they break up into two components:
a left-mover on $\RSC_h$ which is also a left-mover on
the remaining two dimensions $S^1\times\R$,
and a right-mover on $\RSC_h$ which is also a right-mover on
the remaining two dimensions $S^1\times\R.$
Altogether, therefore, the supercharges transform
as a section of the following vector bundle [where the subscript
indicates whether it is a left-mover ($+$) or right-mover ($-$)
on $S^1\times\R$]:
$$
\Bigl\lbrack
\CanK^{\frac{1}{2}-\sum_1^3\twx_i}\oplus
\bigoplus_{i=1}^3\CanK^{\frac{1}{2}+\twx_i}\Bigr\rbrack_{+}
\oplus
\Bigl\lbrack
\CanK^{\frac{1}{2}+\sum_1^3\twx_i}\oplus
\bigoplus_{i=1}^3\CanK^{\frac{1}{2}-\twx_i}\Bigr\rbrack_{-}
\,.
$$
At the same time, scalar fields transform as sections of
\be\label{eqn:scalarsK}
\CanK^{\twx_1+\twx_2}\oplus
\CanK^{\twx_1+\twx_3}\oplus
\CanK^{\twx_2+\twx_3}
\,,
\ee
and their complex conjugates, of course, transform as sections of
\be\label{eqn:scalarsKbar}
\bCanK^{\twx_1+\twx_2}\oplus
\bCanK^{\twx_1+\twx_3}\oplus
\bCanK^{\twx_2+\twx_3}
\,.
\ee

The maximum number of supersymmetry generators that
can be preserved is $4.$
For this  we take $\twx_1=\twx_2=-\twx_3=\tfrac{1}{2}$,
i.e.,
\be\label{eqn:twsu}
\twsu\equiv
\begin{pmatrix}
\frac{1}{2} & &  & \\
& \frac{1}{2} &  & \\
& & -\frac{1}{2}  & \\
& & & -\frac{1}{2}  \\
\end{pmatrix} \in \frak{u}(3)\subset \frak{su}(4)_R\,.
\ee
This is the A-twist discussed in
\cite{Bershadsky:1995vm}\cite{Kapustin:2006pk}.
We see from \eqref{eqn:scalarsK}-\eqref{eqn:scalarsKbar}
that the $6$ real scalars of \SUSY{4} SYM have turned into
$4$ scalars and a $1$-form on $\RSC_h.$

We thus end up with the following setting:
\SUSY{4} SYM compactified on $S^1_R\times\RSC_h$
with an R-symmetry
twist $\gtw$ and an $\SL(2,\Z)$-twist $\zs$ along $S^1_R$ as throughout this paper,
and with an additional A-twist along $\RSC_h.$
We wish to find the Witten index, i.e., the number of supersymmetric vacua
counted with $(\pm)$ signs according to their fermion numbers.

\subsection{The Witten Index}
\label{subsec:WIndex}

The Witten index
is generally independent of continuous parameters,
and as is standard in the computation of a Witten index,
when we identify a useful parameter that
can be taken to different extreme values
we get two opposite limits in which it is interesting to perform the calculation.
In our case, one limit is that the Riemann surface $\RSC_h$ is 
much larger than the circle $S^1_R.$
We refer to it as ``Limit (i).''
In this case, we first reduce to Chern--Simons theory on $\RSC_h\times \R$,
according to our conjecture in \secref{subsec:Conjecture},
and viewing $\R$ as time direction, the Witten index is
just the dimension of the Hilbert space
of Chern--Simons theory.

The method for
calculating the dimension $d_h(n,\lvk)$
of the Hilbert space of $SU(n)$ Chern--Simons theory
at level $\lvk$ on a Riemann surface of genus $h$
was outlined in \cite{Witten:1988hf}. 
The Hilbert space can be obtained by geometric quantization
of the moduli space $\MFC$ of flat $SU(n)$
connections on $\RSC_h$ with a symplectic form
that is $\lvk$ times the K\"ahler $2$-form of $\MFC$, 
which is determined by the complex structure of $\RSC_h$.
Explicit expressions for $n=2$
can be found in \cite{Verlinde:1989hv}.
However, as we will see later, 
we need to modify these equations to include
a nonzero magnetic flux through $\RSC_h$,
and we present the calculation in \appref{app:CS}.

The opposite limit
is to take $\RSC_h$ to be much smaller than $S^1_R.$
We refer to it as ``Limit (ii).''
We can then first reduce \SUSY{4} SYM on $\RSC_h.$
This is precisely the setting studied in
\cite{Bershadsky:1995vm}\cite{Harvey:1995tg}\cite{Kapustin:2006pk}.
With a nonzero magnetic flux on $\RSC_h$, the resulting low-energy description
is a 1+1D $\sigma$-model with a smooth hyper-K\"ahler
target space that can be identified with Hitchin's moduli space $\MH.$
(We will review Hitchin's space in \secref{subsec:HSpace}.)
The magnetic flux is required to make the associated Hitchin space
$\MH$ nonsingular.
S-duality, according to
\cite{Bershadsky:1995vm}\cite{Harvey:1995tg},
reduces to T-duality of the $\sigma$-model.

To compute the Witten index, we compactify the $\R$ direction
on a circle of radius $T.$
(The resulting setting is depicted in \figref{fig:RSxS1xS1}.)
We take periodic boundary conditions along the $S^1_T$
direction for all the $\sigma$-model fields,
and calculate the index in the limit $R\ll T.$
In this limit it is convenient to switch the roles
of time and space and think of $S^1_R$ 
as (Euclidean) time.
The Witten index is then given by the trace
of the T-duality operator $\TKer(\zs)$,
which is the reduction of the $\SL(2,\Z)$ twist $\zs$ to
the Hilbert space of the $\sigma$-model compactified
on $S^1$, times the R-symmetry operator $\gtw$,
treated as an operator in the same Hilbert space
(we hope the reader will forgive this slight abuse
of notation):
\be
I = \tr_{0}\{(-1)^F\TKer(\zs)\gtw\}\,.
\label{modind}
\ee
Here $F$ is the fermion number, and
$\tr_{0}$ denotes the restriction of the trace
to the ground states.

The ground states form a finite dimensional Hilbert
space which can be identified with the cohomology
of the target space $\MH.$
In fact, since the two dimensional space on which
the $\sigma$-model is defined is flat,
we can topologically twist the $\sigma$-model
to get an A-model or B-model \cite{Witten:1991zz} with the same
target space. 
There is a particular complex structure on $\MH$
for which the A- and B-models are invariant under S-duality
(called ``complex structure $I$'' in \cite{Kapustin:2006pk}).
But in any case,
since $\TKer(\zs)\gtw$ preserves supersymmetry
it commutes with the BRST charge, and hence acts
on the finite dimensional Hilbert space of the
topologically twisted theory.
This Hilbert space is identified with the
(de Rham or Dolbeault) cohomology of $\MH$,
and in order to complete the computation of the Witten
index we need to know how $\gtw$ and $\TKer(\zs)$ act
on the cohomology.

\subsection{Review of Hitchin's space}
\label{subsec:HSpace}
It is now time to
review some relevant facts about Hitchin's moduli space
$\MH=\MH(\RSC_h,G)$
associated with a Riemann surface $\RSC_h$ and a gauge
group $G.$
What follows is a list of facts that are relevant to our
discussion, collected from 
\cite{Bershadsky:1995vm,Kapustin:2006pk,
Hitchin:1986vp,HauselThaddeus1-2}.

The Hitchin moduli space is the moduli space of solutions
to the following differential equations:
\be\label{eqn:Fzbz}
F_{z\bz} = [\phi_z,\bphi_\bz]\,,
\qquad
D_z\bphi_\bz = D_\bz\phi_z = 0\,,
\ee
where solutions that are equivalent up to a gauge transformation
are identified in the moduli space. Here $F_{z\bz}$ is the field 
strength of a gauge field with gauge group $G$ on $\RSC_h$, 
$\phi_z dz$ is a $(1,0)$-form 
which takes values in the complexified Lie algebra of $G$, 
$\bphi_\bz d\bz$ is its complex conjugate, and
$D_z\equiv\px{z}-A_z$ and $D_\bz\equiv\px{\bz}-A_\bz$
are the $(1,0)$ and $(0,1)$ parts of the covariant derivative.
(Here, $A_\bz = -A_z^\dagger.$)
We focus on the case with $G=SU(2)$ and assume that the genus 
$h$ of $\RSC_h$ is greater than $1.$

The moduli space $\MH$ in general contains singularities; 
these points correspond to reducible solutions of Hitchin's equation. 
In physical terms, this means that the low energy description 
of \SUSY{4} SYM in terms of $\sigma$-model breaks down 
at these points due to the presence of massless modes 
associated with the residual gauge theory. 
The problem was circumvented in \cite{Bershadsky:1995vm} 
by turning on a nontrivial 't Hooft magnetic flux through $\RSC_h$. 
In fact, one of the main results of \cite{Hitchin:1986vp} was 
that the moduli space $\MH$ becomes a smooth manifold of 
dimension $12h-12$ in this case. Therefore, from now on, 
we will concentrate 
on the moduli space of solutions with magnetic flux turned on.

\subsubsection{Hitchin's fibration}
The crucial point in understanding the T-duality of the
$\sigma$-model with target space $\MH$ is what is called {\it
Hitchin's first fibration} in \cite{Kapustin:2006pk}. 
In this fibration, the base space $B$ is simply parameterized 
by the gauge-invariant polynomials in $\phi_z$; for $G=SU(2)$, 
this is just $b_{zz}=\tr\phi_z^2$, which is holomorphic 
due to Hitchin's equations \eqref{eqn:Fzbz}, 
and hence belongs to $H^0(\RSC_h,\CanK^2)\approx\C^{3h-3}$, 
where $\CanK$ is the canonical bundle on $\RSC_h$. 
The projection map of the fibration simply sends the pair 
$(A,\phi_z dz)$ to $b_{zz}=\tr\phi_z^2$.

At a generic point on the base space $H^0(\RSC_h,\CanK^2)$, 
the holomorphic differential $b_{zz}$ has simple zeroes on $\RSC_h$. 
To obtain the fiber space over this point, 
one first constructs a double cover $\hat{\RSC}_h$ of $\RSC_h$, 
determined by the two-valued differential $\sqrt{b_{zz}}$. 
It is shown in \cite{Hitchin:1986vp} (see also \cite{Bershadsky:1995vm}) 
that the fiber over $b_{zz}$ is then the Prym variety of the double cover 
$\hat{\RSC}_h$. 
(Roughly speaking, this is the space of allowed values of $U(1)$ holonomies, 
where the $U(1)\subset SU(2)$ is determined by the values of $\phi_z$ 
away from the branch points of the double cover.) 
In particular, the fiber is a complex torus with (complex) dimension $3h-3$.

\subsubsection{The most singular fiber}
\label{subsubsec:singf}
While the generic fiber of Hitchin's fibration is
$T^{6h-6}$, there are singular fibers as well at special
values of the holomorphic
quadratic differential $b_{zz}=\tr(\phi_z^2).$
The most singular fiber is over the base point
where the quadratic differential is identically zero: $b_{zz}=0$.
This implies that up to an $SU(2)$ gauge transformation
$\phi_z$ takes the form
\be\label{eqn:phizup}
\phi_z = \left(\begin{array}{cc}
0 & \alpha_z \\
0 & 0 \\
\end{array}\right)\,.
\ee

A special case is when $\phi_z=0$ identically.
The solution to Hitchin's equations then reduces
to finding a flat connection. Thus $\MFC$, the moduli space
of flat connections (for a given magnetic flux), is
a subset of the fiber over $b_{zz}=0.$
The space $\MFC$ is of dimension $6h-6$, so it has the same dimension
as the fiber.

If $\phi_z$ is not identically zero,
then from \eqref{eqn:Fzbz} and \eqref{eqn:phizup}
it is easy to check
that the gauge field must take the form:
\be\label{eqn:Abz}
A_\bz = \left(\begin{array}{rr}
a_\bz & c_\bz \\
0 & -a_\bz \\
\end{array}\right)\,,
\ee
where
$$
a_\bz = -\frac{1}{2}\px{\bz}\log\alpha_z\,,
$$
and $c_\bz$ is arbitrary.
The equation $F_{z\bz}=[\phi_z,\bphi_\bz]$
implies that $c_\bz^*/\alpha_z$ is holomorphic,
and that $\px{z}a_\bz-\px{\bz}a_z = |\alpha_z|^2 +|c_\bz|^2.$

A special case of this is when $c_\bz=0$ identically.
In what follows,
we will only need the case of genus $h=2$ and
with one unit of magnetic flux on $\RSC_2$.
It can then be shown (see \S7 of \cite{Hitchin:1986vp})
that $\alpha_z$ has a 
single simple zero on $\RSC_2$, and the location of this zero
uniquely determines $\alpha_z$ up to a gauge transformation
in $SO(3)=SU(2)/\Z_2.$
($\alpha_z$ is not locally holomorphic, but can be written
as a product of a section of a holomorphic line bundle times
a nonzero function.)
There is an extra complication here due to the
center $\Z_2$ of the gauge group.
If we identify solutions up to {\it any} gauge transformation
in $SO(3)$, including large gauge transformations,
then the space of solutions with $c_\bz=0$ can be identified
with a copy of $\RSC_h.$ (The map from the moduli space
of solutions with $c_\bz=0$ to $\RSC_h$
is given by the location of the zero of $\alpha_z.$)
But if we identify solutions only up to gauge transformations
in $SU(2)$, we have to take into account the existence
of $2^{2h}=16$ classes of large gauge transformations.
Each class is characterized by a map $\pi_1(\RSC_2)\rightarrow\Z_2$
which adds $(\pm)$ signs to the holonomies of the abelian
gauge field $a_\bz d\bz + a_z dz$ along one-cycles of $\RSC_2.$
In this case the space of solutions is a $16$-fold cover
of $\RSC_2$, which is a Riemann surface of genus $17.$
This extra complication will not be important for us,
as we will need only the sector with zero electric flux
along one-cycles of $\RSC_2$, and so for all intents and purposes
of this paper, the space of solutions with $c_\bz=0$
is identified with $\RSC_2.$
(For more details see \S7 of \cite{Hitchin:1986vp}.)

To obtain more information on the singular fiber, 
consider the (real) ``Morse function,'' introduced by Hitchin, on $\MH$:
$$
\mu\equiv 2\int\tr(\phi_z\bphi_\bz) d^2z.
$$
The integral is over $\RSC_h.$
Its minimum is $\mu=0$ and the minimum locus $\mu^{-1}(0)$
is identified with the subspace $\MFC$ of the singular fiber.
For $h=2$, the range of $\mu$ on the singular fiber is
$0\le\mu\le \tfrac{\pi}{2}$, and the maximal value $\tfrac{\pi}{2}$ is attained
on the subspace of solutions with $c_\bz=0.$
We will therefore refer to this subspace as $\mu^{-1}(\tfrac{\pi}{2})$,
and as we have just seen, it is isomorphic to a copy of $\RSC_2.$

Thus, the part of the singular fiber that is not contained
in $\MFC$ is the subset on which $\mu$ takes nonzero values,
i.e., $\mu^{-1}((0,\tfrac{\pi}{2}])$, where $(0,\tfrac{\pi}{2}]$ 
is the set of values
$0<\mu\le \tfrac{\pi}{2}.$
For genus $h=2$,
the set $\mu^{-1}((0,\tfrac{\pi}{2}])$ is an open manifold of (real)
dimension $6$, while $\MFC=\mu^{-1}(0)$ is a closed manifold,
also of (real) dimension $6.$
The boundary of $\mu^{-1}((0,\tfrac{\pi}{2}])$ is a Riemann surface
that is a subset of $\MFC$ and is isomorphic to 
the Riemann surface $\mu^{-1}(\tfrac{\pi}{2})$ (i.e., is $\RSC_2$
if we ignore the $2^{4}$ multiplicity).
To see this note that the boundary of 
$\mu^{-1}((0,\tfrac{\pi}{2}])$ is obtained by setting $\alpha_z=0$
in \eqref{eqn:phizup}, but keeping the upper triangular form 
\eqref{eqn:Abz} for the gauge field.
Define the complex conjugate field $c_z=c_\bz^*.$
Then, Hitchin's equations \eqref{eqn:Fzbz} reduce to
$$
a_\bz = -\frac{1}{2}\px{\bz}\log c_z\,,
\qquad
\px{z}a_\bz-\px{\bz}a_z = |c_z|^2\,.
$$
But these are the same equations that $a_\bz$ and
$\alpha_z$ satisfy on $\mu^{-1}(\tfrac{\pi}{2})$, only that the role
of $\alpha_z$ is played by $c_z.$
Thus, the boundary of $\mu^{-1}((0,\tfrac{\pi}{2}])$, as $\mu\rightarrow 0$,
is isomorphic to $\mu^{-1}(\tfrac{\pi}{2}).$

\subsubsection{Cohomology}
\label{subsubsec:CohomologyMH}
In what follows, we will also need some facts about
the cohomology $H^*(\MH).$
We will restrict to the case of gauge group $SU(2)$
and genus $h=2.$

The Poincar\'e polynomial
\be\label{eqn:Pt}
P(t)\equiv \sum_i \dim H^i(\MH) t^i
\ee
was calculated in \cite{Hitchin:1986vp},
and is given by
\be\label{eqn:Pt2}
P(t) = 1 + t^2 + 4 t^3 + t^4 + t^6
+t^4(1 + 34 t + t^2).
\ee
Let us review how this expression comes about.
The piece $1 + t^2 + 4 t^3 + t^4 + t^6$ is the contribution
of forms supported at $\mu^{-1}(0)$,
and the piece $t^4(1 + 34 t + t^2)$ is the contribution
of forms supported at $\mu^{-1}(\tfrac{\pi}{2})$
(using the notation from \secref{subsubsec:singf}).
Thus, the polynomial $1 + t^2 + 4 t^3 + t^4 + t^6$
is the Poincar\'e polynomial of the moduli
space $\MFC$ of flat $SU(2)$ connections over a genus $h=2$
Riemann surface with one unit of magnetic flux.

The cohomology of $\MFC$ for $SU(2)$ with one unit of magnetic
flux and genus $h>1$
has been calculated in
\cite{Newstead1} (see also \cite{Witten:1992xu}).
It has $2h+2$ generators:
$\alpha\in H^2(\MFC)$,
$\beta_1,\dots,\beta_{2h}\in H^3(\MFC),$
and $\gamma\in H^4(\MFC).$
Let us briefly review where these generators come from.
{}From a flat connection over $\RSC_h$ one can construct
a holomorphic rank-$2$ vector bundle over $\RSC_h.$
Since this vector bundle varies as a function of the point
in $\MFC$ we get a vector bundle over $\RSC_h\times\MFC.$
The second Chern class $c_2$, which is in $H^4(\RSC_h\times\MFC)$
can be decomposed in terms of a basis of $H^*(\RSC_h).$
The coefficient of the generator of $H^0(\RSC_h)$ is $\gamma$,
the coefficient of the generator of $H^2(\RSC_h)$ is $\alpha$,
and the coefficients of the $2h$ generators of $H^1(\RSC_h)$
are the $\beta_j$'s.
There is a quite complicated set of relations \cite{Newstead2}
among $\alpha,\beta_1,\dots,\beta_{2h},\gamma,$
which we will not need in the present paper.
The case of genus $h=2$ is particularly easy to describe.
In this case, by Hodge duality we can complete
the Poincar\'e polynomial of $\MFC$ to
$1+t^2+4t^3+t^4+t^6.$

Another useful fact is that the mapping class group
of $\RSC_h$ acts nontrivially on the generators
$\beta_1,\dots,\beta_{2h}.$
The mapping class group is the fundamental group
of the space of complex structures of $\RSC_h.$
As one traverses a loop in this space,
the complex structure of $\RSC_h$ varies
and with it the space $\MFC$ varies.
As one completes the loop, the complex structure
of $\RSC_h$ is back to its original value,
and the space $\MFC$ is also isomorphic to the
original space at the start of the loop,
but a particular generator of $H^3(\MFC)$ does
not necessarily map to itself.
In general, there is a nontrivial action
[described by an element
in the symplectic group $\Sp(2h,\Z)$ which
preserves the intersection form] on the generators
of $H^1(\RSC_h)$, which induces a nontrivial dual action
on $\beta_1,\dots,\beta_{2h}$.

The remainder of the Poincar\'e polynomial \eqref{eqn:Pt2}
is the contribution from reducible solutions of the form
\eqref{eqn:phizup} with $\alpha_z\neq 0.$
The cohomology that we need is the subspace
invariant under large gauge transformations,
as discussed at the end of
\secref{subsubsec:singf},
and the corresponding Poincar\'e polynomial is
\be\label{eqn:Pt3}
P(t) = 1 + t^2 + 4 t^3 + t^4 + t^6
+t^4(1 + 4 t + t^2).
\ee

The space $\MH$ is noncompact, so we need to specify
whether we allow forms with noncompact support.
Since these correspond to nonnormalizable states,
we will drop them, and so we work with the cohomology
with compact support.
Let us denote by $\delta(\mu^{-1}(0))$ the $6$-form
with support on $\mu^{-1}(0)$ and ``indices'' in the direction
transverse to $\mu^{-1}(0)$ which are the directions of the base
of the Hitchin fibration.
[$\delta(\mu^{-1}(0))$ can be smeared out to what is known 
as the {\it Thom class} of a tubular neighborhood
of $\mu^{-1}(0).$]
Let us also denote by $\delta(\mu^{-1}(\tfrac{\pi}{2}))$ the $10$-form
with support on $\mu^{-1}(\tfrac{\pi}{2})$ (which is a Riemann surface
and therefore has $10$ orthogonal directions).
For the cohomology with compact support we have to multiply
the piece $1 + t^2 + 4 t^3 + t^4 + t^6$ in \eqref{eqn:Pt3}
by $t^6.$ We get representatives
of the cohomology on $\MH$ 
by multiplying the corresponding forms on $\MFC$ by 
$\delta(\mu^{-1}(0)).$
Similarly, $\mu^{-1}(\tfrac{\pi}{2})$ is a Riemann surface
and has a Poincar\'e polynomial $1+4t+t^2$
(ignoring the complication of the $16$-fold cover
mentioned at the end of \secref{subsubsec:singf}).
We get the corresponding forms on $\MH$ by multiplying
the forms on $\mu^{-1}(\tfrac{\pi}{2})$ by $\delta(\mu^{-1}(\tfrac{\pi}{2})).$

\subsubsection{Action of $\gtw$}
\label{subsubsec:ActionOfgtw}
The R-symmetry twist $\gtw$ acts on $\phi_z$ as
$$
\phi_z\rightarrow e^{i\pht}\phi_z,
$$
according to \eqref{eqn:gtwN=6}.
It therefore acts on the quadratic differential as
$$
b_{zz}\rightarrow e^{2i\pht}b_{zz}\,.
$$
Note that $e^{2i\pht}\neq 1$
for all the values of $\pht$ listed in
\secref{subsec:Stwist}.
It follows that a $\gtw$-invariant point of $\MH$
is possible only if $b_{zz}=0.$
Thus, the only fixed points of $\gtw$ occur
over the singular fiber of the Hitchin fibration.
This conclusion holds for any of the values of $n$
and $\pht$ from \secref{subsec:ZModes}.

Restricting to the singular fiber over $b_{zz}=0$,
the $\gtw$-invariant subspace is the disjoint union
$\mu^{-1}(0)\cup\mu^{-1}(\tfrac{\pi}{2})$,
i.e., the union of the moduli space of flat connections
$\MFC$ and solutions with an abelian gauge field ($c_\bz=0$),
which is a copy of $\RSC_2$ (see the notation at the end
of \secref{subsubsec:singf}).
To see this, note that $\phi_z=0$ is obviously $\gtw$-invariant.
This gives $\mu^{-1}(0)$. For nonzero
$\phi_z$ of the form \eqref{eqn:phizup} we have
$$
e^{i\pht}\phi_z =
\left(\begin{array}{cc}
e^{\frac{1}{2}i\pht} & 0 \\
0 & e^{-\frac{1}{2}i\pht} \\
\end{array}\right)
\phi_z
\left(\begin{array}{cc}
e^{\frac{1}{2}i\pht} & 0 \\
0 & e^{-\frac{1}{2}i\pht} \\
\end{array}\right)^{-1}\,.
$$
This gauge transformation, however, doesn't preserve $c_\bz$,
and only $c_\bz=0$ solutions, i.e., those in $\mu^{-1}(\tfrac{\pi}{2})$ 
are $\gtw$-invariant.

\subsubsection{S-duality}
\label{subsubsec:SMH}
As shown in \cite{Bershadsky:1995vm}, the coupling constant of
the four-dimensional gauge theory determines the K\"{a}hler
structure of $\MH$ upon compactification on $\RSC_h$, which is
also the K\"{a}hler structure of each fiber. S-duality in
four dimensions therefore becomes the fiberwise T-duality of
the two-dimensional $\sigma$-model. 

We need to understand the action of $\SL(2,\Z)$-duality on
the A-model operators, i.e., on the de Rham cohomology
of $\MH.$ 
In this subsection we may assume that the $\sigma$-model
is formulated on $\R^2.$
In principle, the action of S-duality 
on a generic fiber of the Hitchin fibration is tractable,
as it reduces to T-duality on the $T^{6h-6}$ fiber.
However, it is not so clear how to track this action
to the singular fiber, which is what we need.
We will thus employ a few indirect arguments.
We will also restrict ourselves to the case $h=2$ and
only the S-duality element $\tau\rightarrow -1/\tau.$ 
We will thus keep denoting it by $\SKer$ but restrict to
$\pht=\pi/2.$

In this section it will be more convenient to work 
with the cohomology $H^*(\MH)$ with noncompact support,
rather than the cohomology of forms
with compact support $H_{\text{cpt}}^*(\MH).$
Since the fibers of the Hitchin fibration are compact,
the distinction between compact and noncompact cohomology
only depends on the behavior of the forms as a function
of the base point.
S-duality preserves the base point,
and in a sense acts classically on the base.
We lose no information by working with noncompact forms.

We can then use the following facts:
\begin{enumerate}
\item[(i)]
$\SKer^2$ acts as charge conjugation on the gauge theory.
For $SU(n)$ gauge group charge conjugation acts on the gauge
field as $A\rightarrow -A^t$ and on the Higgs field as
$\phi_z\rightarrow\phi_z^t$, where $(\cdots)^t$ is the transpose
operation. Combining $\SKer$ with the R-symmetry twist $\gtw$
from \eqref{eqn:gtwN=6} (with $\pht=\pi/2$) we find 
that $(\SKer\gtw)^2$ acts as:
$$
(\SKer\gtw)^2:\qquad
A\rightarrow -A^t\,,\qquad
\phi_z\rightarrow -\phi_z^t.
$$
For $\frak{su}(n)$ with $n>2$ the automorphism $x\rightarrow -x^t$
is outer, but for $\frak{su}(2)$ it is an inner automorphism,
as $-x^t=\sigma_2^{-1} x\sigma_2$ where 
$\sigma_2=\left(\begin{array}{ll}
 0 & -i \\ i & 0 \\ \end{array}\right)
\in SU(2).$
Thus, for an $SU(2)$ gauge group, $(\SKer\gtw)^2$ is 
equivalent to a gauge transformation and acts as the identity
on the A-model operators.
It follows that the eigenvalues of $\SKer\gtw$ are $\pm 1.$

\item[(ii)]
Both $\SKer$ and $\gtw$ commute with the mapping class group
$\Sp(2h,\Z)$ of $\RSC_h.$
The mapping class group $\Sp(2h,\Z)$ acts nontrivially
on the cohomology of $\MH.$
The action of $\Sp(2h,\Z)$ is generated by operations that
can be described as follows.
Suppose we cut $\RSC_h$ along a one-cycle (which we
identify with $S^1$) and glue it back with a rotation
by an angle $\theta$ (a Dehn twist), understood as part of
a holomorphic transformation in a local neighborhood of the cut.
This defines a new complex structure on $\RSC_h$, and as we
let $\theta$ vary continuously from $0$ to $2\pi$ we get a loop
in the moduli space of complex structure of $\RSC_h.$
As we traverse the loop, we can follow
what happens to an integral cohomology class
of $\MH$, and after the loop is completed we generally find
that it is not back to itself. In this way we get a nontrivial
action on $H^*(\MH).$ 
We can therefore decompose $H^*(\MH)$
into irreducible representations of $\Sp(2h,\Z),$
and $\SKer\gtw$ has to act as either the identity or
multiplication by $(-1)$ in each irreducible subspace.
(If an irreducible representation of $\Sp(2h,\Z)$ appears
in the decomposition of $H^*(\MH)$
with multiplicity higher then $1$, 
then $\SKer\gtw$ can mix these subspaces, 
but we can always diagonalize
it in the direct sum of these subspaces
and the eigenvalues will be $\pm 1$.)

\item[(iii)]
The A-model is independent of the complex structure
of the target space altogether, and only the K\"ahler class
is important.
Thus, $\SKer\gtw$ is invariant under complex conjugation.
\item[(iv)]
Some of the operators of the A-model are directly related 
to the topological operator 
$\Op=\int F\wedge F$ of \SUSY{4} SYM.
At the selfdual point $\tau=i$ it is not hard to check
that S-duality acts as $\Op\rightarrow -\Op$
(a small $\theta$-angle is mapped to its negative).
In order to understand to which operators of the 
$\sigma$-model this observation is relevant, 
recall \cite{Witten:1991zz} that a local operator
$\Op^{(0)}$
of the A-model has a nonlocal descendant $\Op^{(2)}$
that can be expressed as an integral over all space $\R^2.$
Let $\Op^{(0)}$ be the operator associated with
the cohomology class in $H^2(\MH)$
that descends from $\alpha\in H^2(\MFC)$ (discussed
in \secref{subsubsec:CohomologyMH}).
(Note that $\Op^{(0)}$ does not correspond to a 
compactly supported class, since we haven't multiplied it
yet by the $6$-form $\delta(b_{zz})$, 
but this is unnecessary for the purposes
of understanding the action of S-duality.)
Since $\alpha$ was defined in terms of the
second Chern class $c_2$, by definition $\Op^{(2)}$
is proportional to $\int F\wedge F$ on the entire space.
We conclude that $\SKer\gtw$ acts as $(-1)$ on $\alpha.$
Applying a similar argument to the $\beta_j$'s in $H^3(\MFC)$
and $\g\in H^4(\MFC)$ runs into minor difficulties, 
since the physical $F\wedge F$ can only be reduced
to a $2$-form on $\R^2$ (by integration on $\RSC_h$),
and, geometrically,
the $\sigma$-model induced map can only turn $\beta_j$
and $\g$ into $3$-forms and $4$-forms respectively.
But the A-model descendent $\Op^{(2)}$ that corresponds to,
say, $\g$ is an integral of a $2$-form on $\R^2$ and contains
two fermionic fields of the A-model.

\item[(v)]
Once we have $\SKer\gtw(\alpha)=-\alpha$ 
we can use the cohomology product to obtain
$\SKer\gtw(\alpha^2)=\alpha^2$ 
and $\SKer\gtw(\alpha^3)=-\alpha^3.$
Here it is crucial to work in cohomology with
noncompact support, since the product is known to 
be trivial for the cohomology with compact support
$H^*_{\text{cpt}}(\MH)$ \cite{Hausel:1998ii}.
As for the cohomology with noncompact support,
$\alpha$ is a $2$-form, which must therefore
be proportional to the K\"ahler class of $\MH$ 
(in complex structure $I$), since $H^2(\MH)$ is $1$-dimensional.
Therefore $\alpha^2$ and $\alpha^3$ are a nonzero
$4$-form and $6$-form, respectively.

\item[(vi)]
We can gather extra clues from the assumption
that S-duality acts as a simple T-duality on the 
generic $T^6$ fiber of the Hitchin fibration 
\cite{Bershadsky:1995vm}.
We have seen in \secref{subsec:Kahler}
that T-duality acts as multiplication by $i^{p+q}$ on 
the operators of the A-model that correspond to elements
in the $H^{(p,q)}$ of Dolbeault cohomology.
This was shown for $T^2$, but the result clearly generalizes
to $T^6.$
This operation does not square to the identity,
but recall from \secref{subsubsec:ActionOfgtw}
that the R-symmetry twist $\gtw$ acts nontrivially
on the base of the Hitchin fibration: 
$b_{zz}\rightarrow -b_{zz}.$
Thus, when discussing the action of $\SKer\gtw$ on
a generic fiber, we have to consider both the fiber
at $b_{zz}$ and the fiber at $-b_{zz}$ simultaneously.
Let $F$ be the fiber over $b_{zz}$ and $F'$ be the fiber
over $-b_{zz}.$
Since $b_{zz}$ and $-b_{zz}$ have the same zeroes
over $\RSC_2$, it follows (by definition of the fibers
as the moduli space of flat connections over a Riemann
surface that is the double cover of $\RSC_2$ branched
over the zeroes of $b_{zz}$) that $F$ and $F'$
are naturally isomorphic.
$\SKer\gtw$ interchanges $F$ and $F'$, and takes the block-form:
\be\label{eqn:Sontwofibers}
\left(\begin{array}{cc}
0 & i^{p+q} \\ (-i)^{p+q} & 0 \\
\end{array}\right),
\ee
where the first block of columns or rows refers to
the $H^{(p,q)}(F)$ and the second block refers to
$H^{(p,q)}(F').$
We have replaced $i^{p+q}$ with $(-i)^{p+q}$ in the second
block so as to keep $(\SKer\gtw)^2=1.$
What can we learn from this about the action of $\SKer\gtw$
on $\MH$? The inclusion maps $\imath:F\hookrightarrow\MH$
and $\imath':F'\hookrightarrow\MH$ induce maps on cohomology
$\imath^*:H^*(\MH)\rightarrow H^*(F)$ and
$\imath^{'*}:H^*(\MH)\rightarrow H^*(F').$
And $\SKer\gtw$ commutes with these maps, in the sense
that $\imath^{'*}\circ(\SKer\gtw)=(\SKer\gtw)\circ\imath^*.$
We can identify $H^*(F)\simeq H^*(F')$ and write
$\imath^*\circ(\SKer\gtw)=(\SKer\gtw)\circ\imath^*.$
Thus, if $\lambda\in H^{(p,q)}(\MH)$ we get
$\imath^*\SKer\gtw(\lambda)=(-i)^{p+q}\imath^*(\lambda).$
This doesn't uniquely determine $\SKer\gtw(\lambda)$
since $\imath^*$ might not be injective,
but it gives us partial information.
For example, $\imath^*$ is injective on $H^{(1,1)}(\MH)$
since it maps the K\"ahler class of $\MH$
(in complex structure $I$) to the K\"ahler class of $F.$
So, we again recover the result that $\SKer\gtw$ acts
as $(-i)^{p+q}=-1$ on the $2$-form $\alpha.$
\end{enumerate}

To summarize,
at this point we have the following information
for $h=2$:
\begin{itemize}
\item
On the $1$-dimensional
$H^0(\MH)$ (or $H_{\text{cpt}}^6(\MH)$)
$\SKer\gtw$ acts as $+1$;
\item
On the $1$-dimensional
$H^2(\MH)$ (or $H_{\text{cpt}}^8(\MH)$)
$\SKer\gtw$ acts as $-1$;
\item
On the $4$-dimensional
$H^3(\MH)$ (or $H_{\text{cpt}}^9(\MH)$)
$\SKer\gtw$ has either $4$ eigenvalues of $-1$
or $4$ eigenvalues of $+1.$
\item
On the $2$-dimensional
$H^4(\MH)$ (or $H_{\text{cpt}}^{10}(\MH)$)
$\SKer\gtw$ has one eigenvalue $+1$ and 
the other eigenvalue is either $+1$ or $-1.$
\item
On the $4$-dimensional
$H^5(\MH)$ (or $H_{\text{cpt}}^{11}(\MH)$)
$\SKer\gtw$ has either $4$ eigenvalues of $-1$
or $4$ eigenvalues of $+1.$
\item
On the $2$-dimensional
$H^6(\MH)$ (or $H_{\text{cpt}}^{12}(\MH)$)
$\SKer\gtw$ has one eigenvalue $-1$ and 
the other eigenvalue is either $+1$ or $-1.$
\end{itemize}
Thus, we know the action of $\SKer\gtw$ up to
four undetermined $(\pm)$ signs.

\subsection{Effect of fluxes: sharpening the conjecture}
\label{subsec:EffectOfFluxes}
We still need to explain which gauge bundle to take for
our conjectured low-energy Chern--Simons theory, i.e.,
what is the three-dimensional magnetic flux.
To answer this question
we will now consider the effect of electric and magnetic fluxes 
of the four-dimensional theory on the 
two-dimensional $\sigma$-model. 
This will lead us to a sharpened version of our conjecture.
The reader who is not interested in the details,
but nonetheless trusts the authors, is advised at this point
to skip to the last sentence of this subsection.

Let us first consider the case with $\tau=i$ and $\zs=\zs'$. 
We saw in \secref{subsec:EMfluxes} that in this case, 
among the Hilbert spaces
$\mathcal{H}_{\ef,\mf}$---each associated with a choice of electric and magnetic fluxes $\ef$ and $\mf$---the 
only ones that are invariant under
the S-duality action of $\zs'$ are those with $\ef=\mf$. 
We will therefore focus our attention on these subspaces.

Since the four-dimensional theory is defined on the space
$\MXf=S_R^1\times\RSC_h\times S_T^1$, 
where the first $S_1$ is regarded as Euclidean time, 
we can decompose the fluxes further in
the following way \cite{Kapustin:2006pk}:
\bear
\ef&=&\ef_0+\ef_1\in \Z_2\oplus H^1(\RSC_h,\Z)\,,\\
\mf&=&\mf_0+\mf_1\in \Z_2\oplus H^1(\RSC_h,\Z)\,.
\label{magdec}
\eear
Roughly speaking, $\ef_0$ and $\ef_1$ are electric fluxes 
along $S_T^1$ and a one-cycle of $\RSC_h$, respectively, 
and $\mf_0$ and $\mf_1$ are magnetic fluxes through $\RSC_h$ 
and a two-cycle consisting of $S_R^1$ and a one-cycle of $\RSC_h$.

As we reviewed in the previous section, 
we need nonzero $\mf_0$ in order
to have a smooth moduli space $\MH$. 
Therefore, we also choose nonzero
$\ef_0$ to have an $\zs'$-invariant Hilbert space. 
After compactifying on $\RSC_h$, nonzero $\ef_0$
implies, according to \cite{Kapustin:2006pk}, 
the presence of a flat $B$-field in the sigma model.

This leaves us the freedom of choice of $\ef_1=\mf_1$. 
To interpret these in the two-dimensional terms, 
we need to distinguish between the moduli
spaces $\MH(\RSC_h,G)$ for different gauge groups $G=SU(2)$ 
and $G=SO(3)$.
The moduli space for $SU(2)$ was briefly described 
in \secref{subsec:HSpace}. 
It is also shown in \cite{Hitchin:1986vp} that the space is
simply connected. On the other hand, it possesses a geometric symmetry group $H^1(\RSC_h,\Z_2)$ 
(which acts by changing the holonomies around the
one-cycles by elements of $\Z_2$), 
and upon dividing the space by this symmetry, 
we get the moduli space for $SO(3)$, whose fundamental group
is $H^1(\RSC_h,\Z_2)$.

It is now clear that states of the $\sigma$-model with target space
$\MH(SU(2))$ will carry conserved momenta corresponding to the
symmetry group $H^1(\RSC_h,\Z_2)$, 
while strings of the sigma model with
target space $\MH(SO(3))$ will carry winding numbers valued in its
fundamental group $H^1(\RSC_h,\Z_2)$. 
The quantities $\ef_1$ and $\mf_1$
signify the conserved momentum and winding number of 
the respective $\sigma$-models, and
the fact that they are exchanged under the T-duality is another
manifestation of the fact that S-duality of 
four-dimensional gauge 
theory reduces to T-duality of the 
two-dimensional $\sigma$-model  upon
compactification \cite{Bershadsky:1995vm,Kapustin:2006pk}. 
Furthermore, it is shown in \cite{HauselThaddeus3}
that the two moduli spaces $\MH(SU(2))$ and $\MH(SO(3))$ are indeed mirror pairs.

It is equally clear, however, that we cannot have nonzero
$\ef_1$ and $\mf_1$ at the same time in the two-dimensional sigma model;
they simply correspond to strings living in two different
target spaces. The only way to achieve $\ef_1=\mf_1$ is to set them
both to zero. States with $\ef_1=0$ in the $\sigma$-model with target space 
$\MH(SU(2))$ are invariant under the action of the symmetry group $H^1(\RSC_h,\Z_2)$, 
so they descend to well-defined states after dividing the moduli space by the symmetry group 
to make the target space $\MH(SO(3))$. On the
other hand, strings moving in $\MH(SO(3))$ with zero winding numbers
can be lifted to strings moving in the covering space $\MH(SU(2))$.
\footnote{In fact, a string in $\MH(SO(3))$ can be  lifted to many copies of strings in $\MH(SU(2))$, related to each other by the action of the symmetry group. The precise statement here is that a state of zero winding number of the $\sigma$-model with target space $\MH(SO(3))$ will lift to a unique state of the $\sigma$-model with target space $\MH(SU(2))$ with zero momentum (i.e., invariant under the action of the symmetry group).}
Therefore, the condition $\ef_1=\mf_1=0$ is consistent.

These considerations lead us to the following formulation of our
conjecture. We start with a four-dimensional gauge theory on
$\MXf=S_R^1\times\RSC_h\times S_T^1$, with S-duality and R-symmetry twists inserted at
a point of $S_R^1$. If we focus on the sector with
$\ef_0=\mf_0=1$ and $\ef_1=\mf_1=0$, then in the limit of small
$\RSC_h$ (limit (ii) of \secref{subsec:WIndex}), 
where the theory becomes a $\sigma$-model with target space $\MH(SO(3))$, 
we end up with the sector with zero winding number
of the $\sigma$-model. 
S-duality becomes T-duality of the $\sigma$-model, 
which sends the states with $\mf_1=0$ to states with $\ef_1=0$
of the $\sigma$-model whose target space is now the universal cover
$\MH(SU(2))$. But states with $\ef_1=0$ are precisely those that can
be interpreted as states of the $\sigma$-model with target space
$\MH(SO(3))$, as explained in the previous paragraph, and hence the
action of T-duality is well-defined in this sector.

On the other hand, in the limit where $S_R^1$ shrinks to a
point (limit (i) of \secref{subsec:WIndex}), we conjecture that the theory becomes a Chern--Simons theory on
$\Mf_3=\RSC_h\times S^1_T$. Therefore, the partition function of Chern--Simons theory 
will calculate the trace of the T-duality times an R-symmetry operator on the
Hilbert space of the $\sigma$-model. Now, 
the $SO(3)$-bundles on $\Mf_3$ are classified according to
their ``magnetic fluxes'' $\mf\in H^2(M_3,\Z_2)$, 
which decomposes as $\mf=\mf_0+\mf_1$ in exactly same way as 
\eqref{magdec}. Hence the partition function can be written as
$$
Z = \sum_{\mf_0,\mf_1} Z_{\mf_0,\mf_1}\,.
$$
Comparing the two descriptions in different limits, 
it is natural to conclude that
$$ I_{1,0} = Z_{1,0}\,,$$
where $I_{1,0}$ is the index, defined in \eqref{modind}, 
of the $\sigma$-model with target space $\MH(SO(3))$ 
restricted to the zero winding number sector, 
and $Z_{1,0}$ is the contribution to the Chern--Simons theory
partition function from bundles with $\mf_0=1$ and $\mf_1=0$. 
This is the sharpened version of our conjecture.

For the other two cases where $\tau=e^{\pi i/3}$ and $\zs=\pm\zs''$, 
we saw in \secref{subsec:EMfluxes} that $\zs$-invariance of the Hilbert space
$\mathcal{H}_{\ef,\mf}$ forces us to choose $\ef=\mf=0$.
This means that the target space of the $\sigma$-model that we get 
upon compactification on $\RSC_h$ is a singular one. We could avoid dealing with
singular target space by inserting Wilson/'t Hooft operators as discussed
at the end of \secref{subsec:EMfluxes}. In the opposite limit, 
we would then have to calculate
the expectation values of these line operators in the Chern--Simons theory,
instead of its partition function. We leave these possibilities 
for a future work, and concentrate on $\tau=i$ and $\zs=\zs'$ case for now.

\subsection{Testing the conjecture}
\label{subsec:Test}
We are now ready to compare the two limits of counting vacua.
In limit (i) $\RSC_h$ is large and we first reduce
on $S^1_R.$ By the conjecture of \secref{subsec:Conjecture}
the result is Chern--Simons theory at level $\lvk.$
There are no low-energy fields left
that carry R-charge, so the twist $\gtw$ has no effect.
The partition function is simply $d_h(n,\lvk,\mf_0)$ 
--- the dimension of the Hilbert space
of $SU(n)$ Chern--Simons theory at level $\lvk$ on $\RSC_h$
with the magnetic flux specified by $\mf_0.$
For $SU(2)$ with one unit of magnetic flux $\mf_0=1$,
genus $h=2$,
and level $\lvk=2$ (corresponding to S-duality
$\tau\rightarrow -1/\tau$ and $\tau=i$) we get 
$d_2(2,2,1)=6$ (see \appref{app:CS} for details).

In limit (ii), $\RSC_h$ is small and we first
reduce on it to obtain a supersymmetric $\sigma$-model
with Hitchin's space $\MH$ as the target space.
We then compactify that on $S^1_R$ with a T-duality twist
and an R-symmetry twist $\gtw.$
The latter acts only on the Higgs fields $\phi_z,\bphi_\bz$
in \eqref{eqn:Fzbz}.
We need the (absolute value of the)
supertrace of $\SKer\gtw$ on the Hilbert
space of the A-model compactified on $S^1$ (i.e., $S^1_T$).
For this, we need to know the action of $\SKer\gtw$
on the A-model states, which are in one-to-one correspondence
with the cohomology of $\MH$ with compact support.
As reviewed in \secref{subsec:HSpace},
$\MH$ has a fibration structure with the base $B$
being the moduli space of gauge-invariant polynomials
in $\phi_z.$ The twist $\gtw$ acts on that space,
and by our restrictions on the rank $n$ and the discussion
in \secref{subsec:ZModes} we may assume that the only fixed
point of $\gtw$ in $B$ is the origin of the Hitchin
fibration where $b_{zz}=\tr(\phi_z^2)=0$
(see the discussion above in \secref{subsubsec:ActionOfgtw}
 for more details). This leaves the singular fiber, 
which is compact, and so the partition function is well defined.
This is the reason why we restrict to
the cohomology with compact support.
In fact, we may just as well restrict to elements of cohomology
that are supported on the singular fiber.
Now we can collect the information from
\secref{subsubsec:SMH} and attempt to reproduce 
in limit (ii) the number $6$
that we got in limit (i). We have to calculate the alternating
sum of traces of $\SKer\gtw$ in the subspaces $H^i(\MH).$
Unfortunately, there are several signs that we did not determine
in \secref{subsubsec:SMH}, and so our conjecture that the sum
is $6$ cannot be tested at this point. 
However, it is easy to see that there are several ways to choose
the undetermined signs so as to reproduce the required result $6$, 
so at this point our conjecture cannot be ruled out either.
In principle, given the exact expressions for the 
representatives of the cohomology of $\MH$, it is possible
to calculate the action of $\SKer$ on them using the general
framework of \cite{Strominger:1996it,Hori:2000kt,Aganagic:2002mp},
but this is beyond the scope of this paper.

\subsection{A six-dimensional perspective}
\label{subsec:SixD}
We will now briefly comment on some aspects
of our construction that can be understood better
in terms of the $(2,0)$-theory.
We mentioned the six-dimensional realization of our setting,
in terms of the $(2,0)$-theory compactified on $T^2$, 
in \eqref{eqn:Idx3z}-\eqref{eqn:dsx3z}.
The identification \eqref{eqn:Idx3z} takes care of the
$\SL(2,\Z)$-twist, but the R-symmetry twist needs to be
added as well. The R-symmetry group of the $(2,0)$-theory
is $\Sp(4)$ [the double cover of $SO(5)$].
While the \SUSY{6}
R-symmetry twist \eqref{eqn:gtwN=6} 
cannot be embedded in $\Sp(4)$,
the \SUSY{4} twist \eqref{eqn:gtwN=4} can be,
if $e^{i\ftw_4}=\pm e^{-\frac{1}{2}i\pht}.$

The lift to six-dimensions introduces a new dimensionful
parameter --- the area $\Area$ of $T^2$, which has to be taken
to zero before all other limits (small $R$ or small $\RSC_h$)
are taken.
So far
we considered two different limits: one in which the size of
$\RSC_h$ is large compared to the size $R$ of $S^1$,
and the other in which $S^1$ is large compared to $\RSC_h$.
We now find yet another possibly interesting limit to consider.
In this limit we take the scale $\sqrt{\Area}$
of $T^2$ to be much larger than $R.$
In the limit $R\ll\sqrt{\Area}$ it is more useful to think
about the space given by \eqref{eqn:Idx3z} as a circle fibration
over an orbifold of $T^2/\Z_\pz$ given by $z\sim e^{i\pht} z$
($\pz=4,6,$ or $3$,
according to whether $\pht=\frac{\pi}{2},\frac{\pi}{3}$,
or $\pht=\frac{4\pi}{3}$).
The radius of the $S^1$ fiber is 
$R\pz$ and the structure group is $\Z_\pz.$
The base $T^2/\Z_\pz$ has several orbifold points, which
are solutions of $z=e^{i\pht}z+n+m\tau$ for some $n,m\in\Z$:
\begin{itemize}
\item
For $\pz=4$ (and $\tau=i$) 
there are $3$ fixed points: $z=0$ and $z=(1+i)/2$
both with monodromy $\Z_4$, 
and $z=\frac{1}{2}$ with monodromy $\Z_2$
(since $z=\frac{1}{2}$ 
is not fixed by the rotation $z\rightarrow e^{i\pht}z$
but is fixed by $z\rightarrow e^{2i\pht}z$).
\item For $\pz=6$ (and $\tau=e^{\frac{\pi i}{3}}$)
there are $3$ inequivalent fixed points:
$z=0$ with monodromy $\Z_6$,
$z=\frac{1}{2}+\frac{i}{2\sqrt{3}}$ with monodromy $\Z_3$,
and $z=\frac{1}{2}$ with monodromy $\Z_2.$
\item For $\pz=3$ (and $\tau=e^{\frac{\pi i}{3}}$) there are $3$ inequivalent fixed points:
$z=0$, $z=\frac{1}{2}+\frac{i}{2\sqrt{3}}$ and
$z=\frac{i}{\sqrt{3}}$, all with monodromy $\Z_3$.
\end{itemize}

The effective description as $R\rightarrow 0$ is weakly coupled
4+1D \SUSY{2} SYM,
with coupling constant $\gYM^2=8\pi^2 R\pz$, on
$\R^{2,1}\times (T^2/\Z_\pz).$
The only complication arises from the fixed points of $\Z_\pz$ listed above.
To understand the behavior of the theory near each fixed point we can replace $T^2$ with $\R^2.$
We are thus led to study the following question:
what is the effective low-energy description of the $(2,0)$-theory
formulated on $\R^{2,1}\times [(\C\times S^1)/\Z_\pz]$ where
$[(\C\times S^1)/\Z_\pz]$ is the orbifold of $\C\times S^1$ 
(parameterized by $(z,x_3)$ with $0\le x_3<2\pi R\pz$) 
by $\Z_\pz$ that is generated by (the freely acting)
$(z,x_3)\mapsto (e^{{2\pi i}{\pz}}z, x_3 + 2\pi R)$?
We can also add an R-symmetry transformation $\gtw$ of order $\pz$
to the above action.

The requisite low-energy description should be formulated on
$\R^{2,1}\times (\C/\Z_\pz)$
and should be 4+1D \SUSY{2} SYM away from the origin of $\C.$
The question is
what are the extra (2+1D) modes that are localized at the origin.
This setting is reminiscent of the Melvin background,
and D-branes in this background have been studied extensively
\cite{Hashimoto:2005hy}. 
Furthermore, using the M-theory realization of the $(2,0)$-theory
as the low-energy theory of M5-branes,
and the relation between the geometry discussed above and
the M-theory lift of $(p,q)$ 5-branes \cite{Witten:1997kz},
one can relate the three-dimensional boundary degrees of freedom
at the singular point of $\C/\Z_\pz$ to the boundary
degrees of freedom at the intersection of D3-branes with
$(1,\pz)$ 5-branes, which were recently solved in 
\cite{Gaiotto:2008sd}\cite{Gaiotto:2008ak}.
Part of the answer is Chern--Simons theory at the fractional 
level $\lvk=1/\pz$ \cite{Kitao:1998mf}.
This can be argued by noting that the bulk action contains
a term of the form
$\frac{1}{2\pi \pz}\int F\wedge F$ where 
the integral is over $\R^{2,1}\times C$ and $C\subset\C/\Z_\pz$
is any open path from $z=0$ to $z=\infty.$
This term can be integrated to give a Chern--Simons
term at level $\lvk=\frac{1}{\pz}$ at the origin
(minus a similar term at infinity).
Thus Chern--Simons actions naturally arise at this limit 
$\sqrt{\Area}\gg R$ as well,
although we have to note that the gauge field variables in this limit do not have a direct
relation to the gauge field variables
in the opposite limit $R\gg\sqrt{\Area}.$

%
%
%

\section{Discussion}
\label{sec:Discussion}
We have put forward various arguments that suggest
that a three-dimensional topological structure underlies
S-duality of \SUSY{4} SYM.
By ``structure'' we mean a 
(probably nonlocal) action $\SKer(\lV,\sdlV)$ that depends
on two independent gauge field configurations and
their supersymmetric partners, but is independent
of the metric or coupling constant.
Here $\lV$ is the ``original'' gauge field configuration $A$
together with the superpartners,
and $\sdlV$ is the ``dual'' gauge field configuration $\sdA$
together with its superpartners.
In other words,
we used the Fredholm kernel representation of an operator
in four dimensions, to construct the action
of a field theory in three dimensions.
It is interesting to wonder whether this setting is
related to a more general framework put forward recently
in \cite{Horava:2008jf} whereby
a wavefunction of a quantum field theory
in $d$ dimensions is related to
the action functional of another quantum field theory
in one dimension less.\footnote{
We wish to thank Petr Ho\v{r}ava for suggesting this connection.}
We have also argued that,
with appropriate modifications and restrictions,
$\SKer(\lV,\sdlV)$ leads to a {\it local} topological
field theory when we set $\lV=\sdlV.$
This field theory had a direct, local description
in terms of a twisted circle compactification.
We conjectured that it is a Chern--Simons theory.

In \secref{subsec:WIndex} we studied the Witten
index of the supersymmetric compactification
on a Riemann surface $\RSC_h$ times an $S^1$
with an R-twist and an $\SL(2,\Z)$-twist.
We considered two different limits: one in which the size of
$\RSC_h$ is large compared to the size of $S^1$,
and the other in which $S^1$ is large compared to $\RSC_h$.
When $\RSC_h$ is large, we could use our conjecture
about the relation to Chern--Simons theory to calculate
the number of vacua. When $\RSC_h$ is small, 
we used the topological string theory on Hitchin's space
to calculate the Witten index in the special case of $SU(2)$
gauge group with an appropriate flux,
$\tau\rightarrow -1/\tau$ twist at $\tau=i$, and 
genus $h=2$.
But we fell short of a full comparison, 
because we did not determine several $(\pm)$ signs
in the action of S-duality. 
As we have argued in \secref{subsec:Test}, 
some sign assignments  are consistent with our conjecture,
and some are not. It would obviously be interesting to 
establish these signs and also extend the tests to higher
genus, other gauge groups, and other $\SL(2,\Z)$ elements.

It would also be interesting to understand in more detail
the T-duality (mirror-symmetry) twist and its relation
to geometric quantization,
as discussed in \secref{subsec:Kahler}.
The general question here is what is the low-energy description
of a $\sigma$-model that is selfdual under mirror symmetry
when compactified on $S^1$ with a mirror symmetry twist.
The simple examples in \secref{subsec:Td} suggest
that the answer is related to geometric quantization 
of the target space. A more general problem can involve
a twist by a combination of mirror symmetry and a geometrical
isometry $\gtw.$ It is then interesting to explore
the relation between the 0+1D low-energy description 
(i.e., the low-energy Hilbert space) and geometric quantization
of the $\gtw$-invariant subspace of target space.
In the context of our setting of \secref{sec:Nonabelian},
the selfdual target space is Hitchin's space $\MH$
associated with $\RSC_h$,
and the low-energy 0+1D theory is Chern--Simons theory
compactified on $\RSC_h$, which can be identified
with geometric quantization of $\MFC$ --- the moduli space
of flat connections on $\RSC_h.$
However, $\MFC$ is not quite the $\gtw$-invariant 
subspace of $\MH.$ 
In the case we considered of gauge group $SU(2)$,
the $\gtw$-invariant subspace is actually a disjoint union
of $\MFC$ and a copy of $\RSC_2$ 
(see \secref{subsubsec:ActionOfgtw} above).

Returning to the general case of a selfdual $\sigma$-model,
the T-duality (mirror symmetry)
twist treats differently the left and right moving
modes of the $\sigma$-model.
One would therefore like to analyze separately the left and right
moving CFTs with the twist.
One tool that might prove useful in this analysis is the recent
construction of Frenkel, Losev and Nekrasov
\cite{Frenkel:2007ux},
where the complex structure $\tau$ of \SUSY{4} SYM
is treated independently from its complex conjugate
$\btau$,
and the limit $\btau\rightarrow\infty$
then reduces the theory
to a simpler topological theory.
For other recent developments in geometric quantization
and its connection to topological string theory, see
\cite{Gukov:2008ve}.

Even if our conjecture about the correspondence
between Chern--Simons theory and
the low-energy limit of the S-duality and R-symmetry
twisted $S^1$ compactification of \SUSY{4} SYM
turns out to be wrong, 
it would still perhaps be interesting to explore
the three-dimensional topological theory that the
twisted compactification defines.
There are quite a few topological quantities that
can be defined through this setting.
For example,
we can study compactification on a Riemann surface
with electric and magnetic fluxes other than selfdual
combinations listed in \secref{subsec:EMfluxes}.
The mismatch between the fluxes of the original
and the dual theories then needs to be corrected
by inserting  Wilson and/or 't Hooft line operators,
as outlined at the end of \secref{subsec:EMfluxes}.
Even on $\R^3$,
the twisted \SUSY{4} compactification
also defines expectation values for knots,
which we expect to be topological,
at least for the limited list of ranks and twists
listed in \secref{subsec:Conjecture}.
It would perhaps be interesting to check if these
knot invariants agree with those calculated from
Chern--Simons theory, or if they give rise to different
knot invariants. For recent developments on the connection
between knot invariants and string theory, see 
\cite{Gukov:2007tf,Gukov:2007ck}.
More generally, it would perhaps be interesting to study
the partition function of the twisted compactification
on $\Mf_3\times S^1$, where $\Mf_3$ is a general $3$-manifold.
It would also be interesting to extend the discussion
to selfdual theories with less supersymmetry.
It was recently shown that a Chern-Simons term 
in three-dimensions can be induced
in certain circle compactifactions
of chiral four dimensional theories with a flavor symmetry twist
\cite{Poppitz:2008hr}. It would perhaps be interesting
to generalize this to include a duality twist.

In order to get a topological low-energy theory
we had to restrict the rank of the $SU(n)$ gauge group
to $n\le 5$ (see \secref{subsec:Conjecture}).
For higher values of $n$ (and even for lower values,
for some of the  $\SL(2,\Z)$ twists) we get 
scalar and spinor zero modes, and the low-energy description
is not topological. Nonetheless, we get in this way nontrivial
three-dimensional theories with \SUSY{6} supersymmetry,
and it would perhaps be interesting to explore their
connection to the superconformal theories
that were recently discovered in connection
with M2-branes at orbifold singularities
and supersymmetrization of Chern--Simons theory
\cite{Schwarz:2004yj}-\cite{Aharony:2008ug}.

More ambitiously, 
we would like to gain new information about S-duality itself.
For this we need to understand the full 
topological structure of the 
S-duality kernel $\SKer(\lV,\sdlV).$
One possible direction might be to start with
the assumption \eqref{eqn:WCC} about the expectation value
of pairs of Wilson loops and attempt to reconstruct
$\SKer(\lV,\sdlV)$ from it. This can be done, in principle,
on a lattice, but it would be perhaps interesting to study
if it has a meaningful continuum limit.
\\[10pt]

{\bf Acknowledgments}
\newline\noindent
We are grateful to Mina Aganagic, Chris Beem, Aaron Bergman,
Eric Gimon, M{\aa}ns Henningson, Petr Ho\v{r}ava, 
Soo-Jong Rey, Mithat \"Unsal,
David Vegh, and Edward Witten for discussions and correspondence.
OJG wishes to thank Rob Myers, Mark van Raamsdonk and Wati Taylor,
the organizers of the workshop on
``Emerging Directions in String Theory''
(where preliminary results of this work have been presented),
and the Banff International Research Station for Mathematical
Innovation and Discovery (BIRS) for their hospitality.
This work was supported in part by the 
Berkeley Center of Theoretical Physics,
in part by the U.S. National Science Foundation
under grant PHY-04-57315,
and in part by the Director, 
Office of Science, Office of High Energy Physics, 
of the U.S. Department of Energy under Contract 
No. DE-AC03-76SF00098.

\appendix
\section{Chern--Simons theory with magnetic flux}
\label{app:CS}
\FIGURE{
\begin{picture}(400,130)

\put(85,70){\begin{picture}(200,100)
\put(-3,15){$\RSC_h$}
\thinlines
\qbezier(0,10)(10,10)(20,20)
\qbezier(20,20)(30,30)(50,30)
\qbezier(50,30)(70,30)(80,20)
\qbezier(80,20)(90,10)(90,0)

\qbezier(0,10)(-10,10)(-20,20)
\qbezier(-20,20)(-30,30)(-50,30)
\qbezier(-50,30)(-70,30)(-80,20)
\qbezier(-80,20)(-90,10)(-90,0)

\qbezier(0,-10)(10,-10)(20,-20)
\qbezier(20,-20)(30,-30)(50,-30)
\qbezier(50,-30)(70,-30)(80,-20)
\qbezier(80,-20)(90,-10)(90,0)

\qbezier(0,-10)(-10,-10)(-20,-20)
\qbezier(-20,-20)(-30,-30)(-50,-30)
\qbezier(-50,-30)(-70,-30)(-80,-20)
\qbezier(-80,-20)(-90,-10)(-90,0)

\qbezier(-75,2)(-65,-8)(-50,-8)
\qbezier(-50,-8)(-35,-8)(-25,2)

\qbezier(-70,0)(-60,10)(-50,10)
\qbezier(-50,10)(-40,10)(-30,0)

\qbezier(75,2)(65,-8)(50,-8)
\qbezier(50,-8)(35,-8)(25,2)

\qbezier(70,0)(60,10)(50,10)
\qbezier(50,10)(40,10)(30,0)
\end{picture}}

\put(185,70){\begin{picture}(200,100)
\put(0,-15){\line(1,0){50}}
\put(0,15){\line(1,0){50}}
\put(65,0){\line(-1,-1){30}}
\put(65,0){\line(-1,1){30}}
\end{picture}}

\put(355,70){\begin{picture}(200,100)
\thinlines
\qbezier(0,10)(10,10)(20,20)
\qbezier(20,20)(30,30)(50,30)
\qbezier(50,30)(70,30)(80,20)
\qbezier(80,20)(90,10)(90,0)

\qbezier(0,10)(-10,10)(-20,20)
\qbezier(-20,20)(-30,30)(-50,30)
\qbezier(-50,30)(-70,30)(-80,20)
\qbezier(-80,20)(-90,10)(-90,0)

\qbezier(0,-10)(10,-10)(20,-20)
\qbezier(20,-20)(30,-30)(50,-30)
\qbezier(50,-30)(70,-30)(80,-20)
\qbezier(80,-20)(90,-10)(90,0)

\qbezier(0,-10)(-10,-10)(-20,-20)
\qbezier(-20,-20)(-30,-30)(-50,-30)
\qbezier(-50,-30)(-70,-30)(-80,-20)
\qbezier(-80,-20)(-90,-10)(-90,0)

\qbezier(-75,2)(-65,-8)(-50,-8)
\qbezier(-50,-8)(-35,-8)(-25,2)

\qbezier(-70,0)(-60,10)(-50,10)
\qbezier(-50,10)(-40,10)(-30,0)

\qbezier(75,2)(65,-8)(50,-8)
\qbezier(50,-8)(35,-8)(25,2)

\qbezier(70,0)(60,10)(50,10)
\qbezier(50,10)(40,10)(30,0)

\thicklines
\qbezier(70,0)(80,20)(90,0)

\thicklines
\qbezier(-70,0)(-80,20)(-90,0)

\thicklines
\qbezier(0,-10)(-5,0)(0,10)

\put(3,-3){$\ket{i}$}
\put(-18,-3){$\bra{i}$}

\put(65,12){$\ket{j}$}
\put(75,-6){$\bra{j}$}

\put(-72,12){$\ket{l}$}
\put(-88,-6){$\bra{l}$}

\end{picture}}

\end{picture}
\caption{
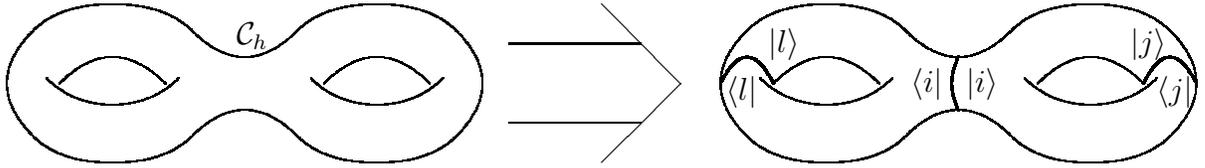
The partition function of Chern--Simons theory
on $\RSC_2$ times $S^1$ (not shown) is calculated as follows
\cite{Witten:1988hf}:
(i) the Riemann surface $\RSC_2$ is cut in three places
along three circles;
(ii) a complete set of states (of the Hilbert space on $T^2$
which is the cut times the $S^1$ that is not shown)
is inserted on the two sides of each cut, 
$\ket{l}\bra{l},\ket{i}\bra{i},\ket{j}\bra{j}$;
(iii) the cuts are 
separated to form two $3$-holed spheres
and the partition function is calculated using the fusion
rules as $\sum_{ijl}N_{lli}N_{ijj}.$
}
\label{fig:Nijl}
}
The dimension of the Hilbert space of $SU(2)$ Chern--Simons
theory at level $\lvk$ on a Riemann surface of genus $h$
is given by \cite{Verlinde:1989hv}:
\be\label{eqn:dh}
d_h(2,\lvk)=
\left(\frac{\lvk}{2}+1\right)^{h-1}
\sum_{j=0}^{\lvk}
\frac{1}{\sin^{2(h-1)}(\frac{j+1}{\lvk+2})\pi}
\,.
\ee
This, by definition, has no magnetic flux.
However, in this paper we need the dimension 
of the Hilbert space of $SU(2)/\Z_2$ gauge configurations
with one unit of magnetic flux
($\mf_0=1$ and $\mf_1=0$ in the notation of 
\secref{subsec:EffectOfFluxes}).
In this appendix we will outline the calculation
of this dimension, following 
\cite{Witten:1988hf}\cite{Verlinde:1989hv}.
We focus on the case of genus $h=2$, but it is easy
to generalize to higher genus.

The dimension $d_h(2,\lvk,\mf)$ is equal to
the partition function on $\RSC_2\times S^1$
with the gauge bundle specified according to the magnetic
flux $\mf.$ We refer to $S^1$ as (Euclidean) ``time.''
The partition function can be calculated by cutting
the Riemann surface along three loops, as in
\figref{fig:Nijl}. Each cut has the topology of a torus
$S^1\times S^1$, where the second factor is time
and the first factor is the appropriate loop
that corresponds to one of the three one-cycles
along which we cut. 
Thus $\RSC_2$ is realized as two $3$-holed spheres glued
appropriately along the holes.
Next, for each cut (hole)
we insert a complete set of states
$\ket{i}\bra{i}$ of the Hilbert space of Chern--Simons theory
on $S^1\times S^1$ and calculate the partition function
using the known expressions for the partition function
of Chern--Simons theory on a $3$-holed sphere with
boundary states $\ket{i_1},\ket{i_2},\ket{i_3}$,
i.e., the ``fusion rules'' $N_{i_1 i_2 i_3}.$
If we insert the complete sets of states
$\ket{l}\bra{l},\ket{i}\bra{i},\ket{j}\bra{j}$
in the left, middle, and right cut, respectively
(see \figref{fig:Nijl}), then the partition function
without magnetic flux
can be written as
as $\sum_{ijl}N_{lli}N_{ijj}.$

To introduce magnetic flux $\mf_0=1$ we can insert a large 
gauge transformation along one of the cuts, say the middle one,
before gluing. The large gauge transformation 
corresponds to a topologically nontrivial map 
$S^1\rightarrow SO(3)$ and if we denote its action
on the states by $\ket{i}\mapsto \sum_{p}\Lambda_{ip}\ket{p}$
we get the partition function
$$
d_2(2,\lvk,\mf_0=1)=\sum_{ipjl}N_{lli}\Lambda_{ip}N_{pjj}.
$$
To get explicit expressions we use the basis of states
on $T^2\simeq S^1\times S^1$ introduced in \cite{Witten:1988hf}.
The states are denoted by $\ket{l}$ with $l=0,\dots,\lvk$
labeling a representation of spin $l/2.$
The state $\ket{l}$ can be realized by filling the first
$S^1$ factor to form a disc, then inserting a closed Wilson line
in the representation with spin $l/2.$ 
If we assume that the Wilson line
runs in the time direction (the second $S^1$ factor)
and is located at, say, the origin of the disc
then the fusion rules have a simple expression
\cite{Verlinde:1988sn}:
$$
N_{i_1 i_2 i_3} = \left\{\begin{array}{ll}
1 & \text{if $|i_2-i_1|\le i_3\le i_1 + i_2$, 
          and $i_1+i_2+i_3\le 2\lvk$;} \\
0 & \text{otherwise.}
\end{array}\right.
$$
On the other hand, if we define the basis of states
by filling the second (time direction) $S^1$ instead,
and let the Wilson lines run parallel to the first $S^1$ factor,
we get a basis of states $\ket{i'}$ on which 
the large gauge transformation is easy to describe
(since it acts on the Wilson line in a simple way):
$$
\Lambda_{i'p'} = \delta_{i'p'}(-1)^{i'}.
$$
The unitary transformation from the basis $\{\ket{j}\}$
to the basis $\{\ket{j'}\}$ is given by \cite{Verlinde:1988sn}:
$$
\ket{j'} = \sum_j S_{j'j}\ket{j},\qquad
S_{j'j} = \sin \frac{(j+1)(j'+1)\pi}{k+2}\,.
$$
Using this expression the partition function can be calculated
as
$$
d_2(2,\lvk,\mf_0=1)=
\sum_l (-1)^l\left(\sum_p S_{lp}\sum_j N_{pjj}\right)^2.
$$
For $\lvk=2$ we get $d_2(2,2,\mf_0=1)=6.$
This agrees with the result of \cite{Daskalopoulos:1993sv}.



\end{document}